\documentclass[authoryear]{elsarticle}

\makeatletter
\def\ps@headings{%
\def\@oddhead{\mbox{}\scriptsize\rightmark \hfil \thepage}%
\def\@evenhead{\scriptsize\thepage \hfil \leftmark\mbox{}}%
\def\@oddfoot{}%
\def\@evenfoot{}}
\makeatother
\usepackage{subcaption}
\usepackage{graphicx}
\usepackage{amsmath,amsfonts,amssymb}
\usepackage{amstext}

\usepackage{hyperref}
\usepackage{comment}
\usepackage{soul} 

\usepackage[inline]{enumitem}
\usepackage{float}
\usepackage{adjustbox}
\usepackage{array, caption, tabularx,  ragged2e,  booktabs}

\usepackage{listings}
\usepackage{longtable}
\newcolumntype{P}[1]{>{\RaggedRight\arraybackslash}p{#1}}
\usepackage{pdflscape}
\usepackage{xcolor}
\usepackage{url}
\usepackage{svg}

\usepackage{booktabs}

\newtheorem{remk}{Remark}

\begin{document}
\begin{frontmatter} 

\title{ActDroid: An active learning framework for Android malware detection}

 \author[hwDu]{Ali Muzaffar\corref{cor1}}
 \ead{am29@hw.ac.uk}
 \author[hwDu]{Hani Ragab Hassen}  
 \ead{h.ragabhassen@hw.ac.uk}
  \author[hwDu]{Hind Zantout}  
 \ead{h.zantout@hw.ac.uk}
   \author[hwEd]{Michael A Lones}  
 \ead{m.lones@hw.ac.uk}
 \cortext[cor1]{Corresponding author}

 \address[hwDu]{Heriot-Watt University, Dubai, UAE}
\address[hwEd]{Heriot-Watt University, Edinburgh EH14 4AS, United Kingdom}

 \begin{abstract}
 The growing popularity of Android requires malware detection systems that can keep up with the pace of new software being released. According to a recent study, a new piece of malware appears online every 12 seconds. To address this, we treat Android malware detection as a streaming data problem and explore the use of active online learning as a means of mitigating the problem of labelling applications in a timely and cost-effective manner. Our resulting framework achieves accuracies of up to 96\%, requires as little of 24\% of the training data to be labelled, and compensates for concept drift that occurs between the release and labelling of an application. We also consider the broader practicalities of online learning within Android malware detection, and systematically explore the trade-offs between using different static, dynamic and hybrid feature sets to classify malware.
 
 \end{abstract}


\end{frontmatter}


\section{Introduction}
The smartphone industry has seen an exponential growth in recent years, with Android emerging as the leading OS provider. For instance, in 2023, over 70\% of surveyed smartphone users were found to use an Android device \citep{androidData}. However, with this growth comes challenges.  Malware is one of these challenges, and is particularly a problem for Android because of the open nature of the platform. As an example of this, it is commonplace for users to download applications from unregulated third-party application repositories where malware is common. To give an idea of the scale of this problem, it has been estimated that a new piece of Android malware is found in online repositories every 12 seconds \citep{malwareStat}. 


Numerous studies have shown that machine learning (ML) models perform well in identifying malware. In two recent reviews of work in this area, we surveyed \citep{Muzaffar2021} and experimentally compared \citep{Muzaffar2023} the merits of different ML modelling approaches for Android malware detection. However, most studies published in this area treat malware detection as a batch learning problem; that is, examples of malware and benign applications are collected, and a single model is then trained to discriminate between these two classes. This approach is unlikely to work well in practice because Android malware changes over time, a process known as \textit{concept drift}. Online learning (OL) is a common approach to dealing with concept drift, and is based on the idea of treating data as a stream and incrementally training an ML model as new data becomes available. A number of previous studies have applied OL to Android malware detection, but we argue that these previous studies did not fully account for the practicalities of implementing OL in a real world context.

In particular, we focus on the challenge of labelling, i.e.\ the process by which applications become labelled as malware or benign. In previous work, it has been assumed that a correct label for an application is available as soon as the application is released. We refer to this scenario as \textit{progressive validation}. However, in practice, there is almost always a delay between application release and the availability of a label. For instance, from our application dataset, we estimate an average delay of 40 days between the release of an application and the generation of a VirusTotal \cite{virustotal} report that can be used to reliably determine its label. We refer to this more realistic situation as \textit{delayed progressive validation} and show that it leads to significant degradation in the accuracy of OL models due to concept drift that occurs between the release of an application and the availability of a label for training. In addition to the delay caused by labelling, another important issue is its cost; for instance, VirusTotal offers only a limited number of API calls for free, meaning that it can be impractical for a system that processes hundreds of applications every day to acquire a label for each application.

In this paper, we address these problems using a novel approach based around the concept of active learning. Rather than training on all available data, active learning models select which data samples they learn from in order to improve their performance. In our framework, this happens in two ways: first, a model only trains on samples for which it has low confidence; second, when a model detects concept drift in its outputs, it retrains from scratch using a set of recent data.
Within this context, this paper makes the following contributions:


\begin{itemize}
    \item We introduce the concept of delayed progressive validation in OL-based Android malware detection, and demonstrate that modelling choices made in the absence of labelling delays can be misleading when models are evaluated within this more realistic scenario.
    \item We introduce the concept of active learning within OL-based Android malware detection and show that this not only compensates for the loss of accuracy due to labelling delays but also allows models to be trained using up to 76\% less labelled data.
    \item In order to better understand the design decisions and trade-offs within OL-based Android malware detection systems, we systematically explore the strengths and weaknesses of different base models and different static, dynamic and hybrid features.
    

\end{itemize}

\section{Related Work}

In this section we first review past works that used ML methods to detect Android malware, and then review works that specifically used online learning.

\subsection{Machine learning in Android malware detection}

Android malware detection based on ML has shown promising results \citep{Muzaffar2021,Muzaffar2023}. The most common approach is to build ML models from features derived using static analysis of the code. Although various features can be extracted in this way, permissions and API calls have been particularly widely used for building malware detection models. Of these, API calls appear to be especially useful for building accurate models \citep{Peiravian2013,Rathore2021,Ma2019}. However, the number of API calls available in recent Android releases is very large, requiring judicious feature selection \citep{Muzaffar2021}. Other authors have found opcodes to be a useful static feature for building accurate models \citep{Bai2020,Kang2016,Xiao2019a}, and the Drebin feature set \citep{Arp2014} has also been used to construct accurate models in past works \citep{Li2017,Wang2017}, though our more recent results \citep{Muzaffar2023} suggest that this combination of features is less relevant for detecting contemporary Android malware.

Fewer works have used features extracted through dynamic analysis of executing programs, largely due to the higher computational cost of carrying out dynamic analysis. Interestingly, API call information captured through dynamic analysis tends to be less useful than static API calls for building models \citep{Muzaffar2023}, possibly due to the difficulty of ensuring code coverage during dynamic analysis. However, several studies \citep{Hou2017,Vinod2019,Xiao2019} report using system calls to build models with high detection rates. Perhaps surprisingly, network-based features have been found to lead to the most accurate models \citep{Zulkifli2018,Wang2016,Muzaffar2023}, with rates up to 99\% reported, although the difficulty of reliably extracting these features may limit their practical use.

Hybrid frameworks, which combine static and dynamic analysis, are the least prominent. \citet{Kandukuru2017} and \citet{Shyong2020} both combined permissions and network traffic features, and reported accuracies of up to 99\%. Given the issues surrounding network-based features, in \citep{Muzaffar2023} we considered other combinations of static and dynamic features, and found that similar levels of accuracy could be achieved by ensembling models that use static API calls and dynamic system calls. Frameworks such as \citet{Lindorfer2015}'s MARVIN and \citet{Saracino2018}'s MADAM have also used market information and other metadata to construct models, leading to detection rates of up to 98\%.

For a recent review of previous studies, see  \citep{Muzaffar2021}. There is considerable variety in the size, age and balance of datasets used for Android malware detection studies, and this makes comparisons based on published results challenging. For more information about the relative importance of different features within a consistent experimental framework, see \citep{Muzaffar2023}, in which we reimplemented 16 representative studies using a large contemporary dataset.


\subsection{Online learning in Android malware detection}

Most ML approaches use batch learning, in which a single model is learnt from a fixed data set. Online learning (OL), on the other hand, trains models incrementally, and is used in situations where the data distribution changes over time, a phenomenon known as concept drift \citep{singh2012}. Concept drift is very likely to occur in malware, since attackers continually exploit new vulnerabilities and try to avoid novel detection techniques, including those which use ML. This means that online learning is likely to be a more suitable basis for building and evaluating malware detection models.


Although used considerably less than offline ML, a number of previous studies have considered the use of OL in Android malware detection to deal with the problem of concept drift. In one of the earliest studies, \citet{Narayanan2016} introduced an OL based framework, called \textit{DroidOL} in which they extracted inter-procedural control-flow graphs and then used a Weisfeiler-Lehman graph kernel to train a passive-aggressive classifier. They reported an accuracy of 84.29\%, measured using a dataset of 44,347 benign and 42,910 malicious applications collected in 2014. A year later, they published an improved framework, \textit{CASANDRA} \citep{Narayanan2017} with a reported accuracy of 89.29\%.

A number of more recent studies have reported higher rates of malware detection. \citet{Mirzaei2019}, in their framework \textit{ANDRODET}, trained leveraging bag, LearnNSE, Hoeffding tree, stochastic gradient descent, weighted majority algorithm and Naïve Bayes online algorithms, and reported accuracies of up to 92.02\%. They  used the AMD \citep{AMD} dataset consisting of 34,962 applications released from 2010 to 2016. \citet{Xu2019}, in their framework \textit{DroidEvolver}, used a dataset of 33,294 benign and 34,722 malicious applications. DroidEvolver is notable for maintaining an application buffer and using pseudo-labels to update the models. Every new application that is classified is checked with the buffer to detect drifting in the models. Xu et al.\ reported an F1-Score of 95.27\%, which declines by 1.06\% on average over five years, though a later study showed that the model can poison itself with this process \citep{kan2021}. \citet{ceschin2023} took a similar approach to DroidEvovler, by updating both the classifiers and features when drifting is detected. Using textual attributes as features, the authors reported accuracies of up to 99\% and 89.66\%, respectively, on two different datasets.

Although past studies have reported high levels of malware discrimination, they have not yet considered the delay between an application's release and the availability of a correct label, or they use pseudo-labels only to update the models, and therefore the results may not accurately represent how the frameworks would work in real-world scenarios and endpoints.  

\section{Experimental Framework}

In this study, we introduce a new active learning OL framework for Android malware detection. We compare it against non-active approaches to OL in both progressive validation and delayed progressive validation scenarios, the former to determine how well it compares against an ideal situation where labels are always available, and the latter to determine its real benefits over non-active approaches. Since different OL models may have different inductive biases, to provide a fair comparison, we use five different base OL models. We then train each of these models separately on static, dynamic and hybrid (i.e.\ a combination of static and dynamic) features. We use the most common static, dynamic and hybrid features in Android malware detection, as reported in \citet{Muzaffar2022,Muzaffar2023}.



\subsection{Models}

We use the following five OL models in our framework:

\paragraph{Passive-aggressive classifier (PA)} The passive-aggressive classifier (PA) can handle large datasets and is one of the most commonly used OL models. PA is a linear model and tries to remain as close to the current model at each iteration as possible. PA uses a regularization parameter to calculate the distance between the current model and the new model and tries to keep the distance close to zero. The classifier remains \textit{passive} if the distance is zero. Otherwise, the classifier updates its weights in order to better classify future instances of data, referred to as being \textit{aggressive} \citep{crammer2006}.

\paragraph{Hoeffding tree} The Hoeffding tree classifier is an online variant of standard decision tree models. Instead of being trained once on a whole dataset, a Hoeffding tree is incrementally updated as new instances arrive, with new splits being introduced as necessary according to a statistical test called the Hoeffding bound, based on a small random sample of data. This allows the tree to split nodes without examining all the data and builds the tree incrementally while adapting to changes. This makes Hoeffding trees more efficient than traditional decision tree approaches while handling large datasets or streams of data \citep{moa}. 

\paragraph{Adaptive random forest} Adaptive random forest combines the strengths of random forest and OL. The adaptive random forest creates an ensemble of decision trees and uses a weighted average to produce the final classification prediction, similar to a traditional random forest. Adaptive random forests, however, update their trees in real-time as new data arrives. Adaptive random forests also monitor the performance of their elements over time and replace a tree if its performance drops \citep{Gomes2017}.  

\paragraph{K-nearest neighbour (kNN)} This is the online variant of the traditional k-nearest neighbour model. The online variant of kNN stores a buffer of recent instances to find the nearest neighbours (using Euclidean distance) to make a final prediction. As the online variant only stores a fixed number of instances, it can make predictions more quickly and efficiently than a traditional kNN \citep{moa}. 

\paragraph{Gaussian Naïve Bayes} Gaussian Naïve Bayes uses probability distributions to predict the class of a new instance by calculating the mean and variance of each feature of each class. Online Gaussian Naïve Bayes updates the model with each new instance and so is able to adapt to changes in data more efficiently than a traditional Gaussian Naïve Bayes model \citep{moa}.

\subsection{Drift Detection}

Using fixed datasets to train models, like in traditional ML, may lead to concept drift. OL adapts to concept drift by learning from data arriving as a stream in real time. Drift detection methods can be used to detect drifting in ML models. One such algorithm is Adaptive Windowing (ADWIN), which monitors the average of a sliding window of the data, and if the average changes significantly over time, it detects change and updates the window size accordingly \citep{bifet2007a}. Specifically, ADWIN compares two sub-windows using a statistical test to identify a significant difference between them. If the distribution equality does not hold any more between the sub-windows, ADWIN tags it as concept drift. We use ADWIN to detect concept drift in this work.

\subsection{Data}
We used a dataset consisting of applications downloaded from various application stores with release dates between 31 December 2018 and 17 April 2021, described in \cite{Muzaffar2023}. In order to recreate a realistic OL process, we used the application release dates, which were saved as metadata. However, this was not possible for malware as those repositories do not provide information related to release dates. There are several ways of estimating a release date of an application, from checking the last edit in the APK file, to the earliest timestamp in the APK file. Timestamps can also be retrieved from Dex files (compiled Android APK code) or the manifest file. However, according to \citet{GUERRAMANZANARES2022}, these methods are not as reliable as using the VirusTotal report date to estimate the release date of the application, which provides information about when the application was first seen online. Therefore, we used the VirusTotal report date as a baseline to determine the release date of malware.

\begin{figure}[t]
\includegraphics[width=\columnwidth]{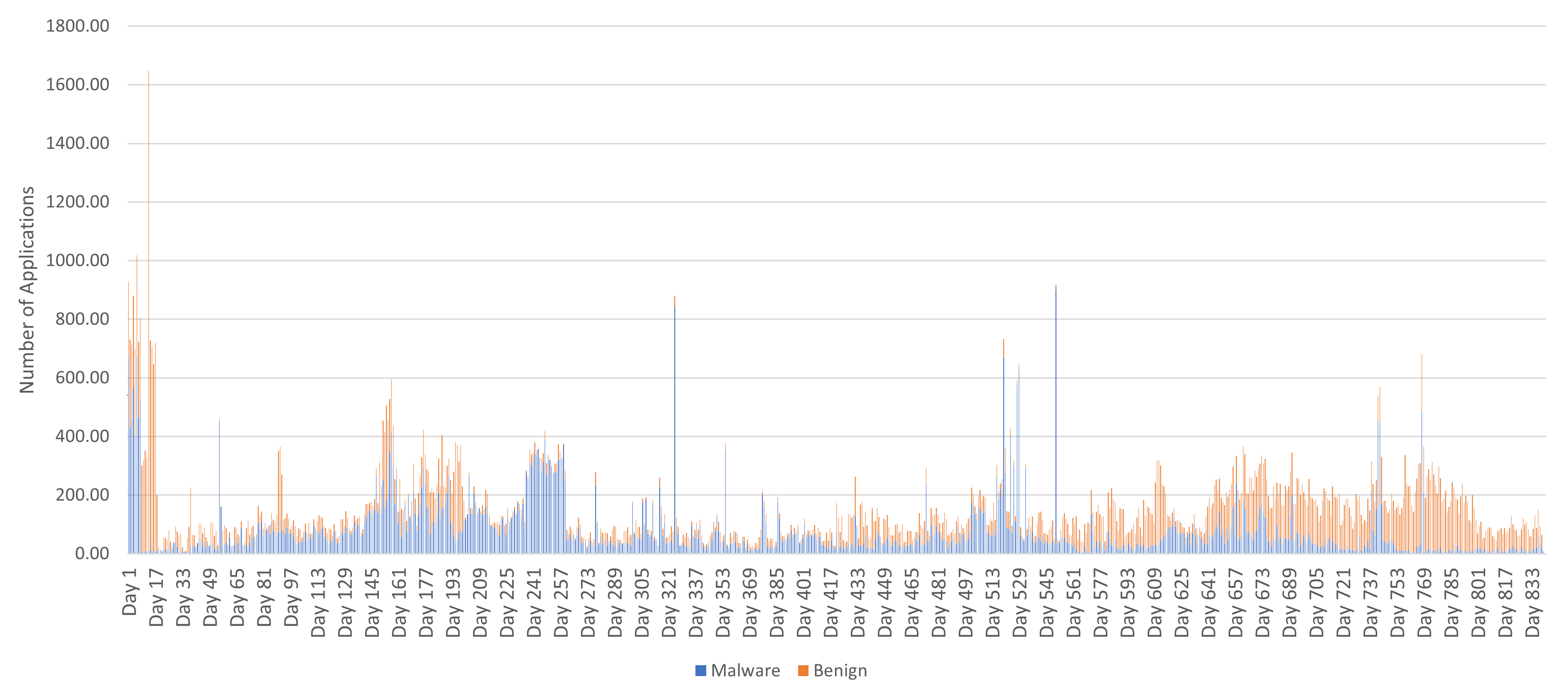}
\centering
\caption{Number of benign and malware applications in the dataset per day}
\label{fig:appol}
\end{figure}

We grouped the applications by date and used them to train OL models. Figure \ref{fig:appol} shows the number of benign and malware applications for each day. We tried to simulate a real-world scenario of downloading the applications from application stores and updating the dataset every day with newly released applications. Hence our data stream is made up of days, and each day consists of applications, benign or malicious, released on that day. In total we used 62,000 benign and 62,000 malware applications for static features. During dynamic analysis some applications terminate unexpectedly; the set of applications for which we were able to complete dynamic analysis was 53,960 benign and 53,202 malware.

\subsection{Features}

We use the following static features:

\begin{itemize}
    \item \textbf{Permissions:} The most commonly used static feature in the literature. The application states the permissions used in the manifest file, making this feature easy to extract. Permissions form the core of the Android security system and are essentially a set of rules that provide applications with access to different parts of the device, including hardware components.
    \item \textbf{API calls:} In offline ML models, API calls have been found to be the best-performing static feature, so it makes sense to also consider them in an OL setting. API calls are requests made by applications to the operating system to perform a specific task. These are used to access system-level functionality. 
    \item \textbf{Opcodes:} Opcodes provide insights into low-level application behaviour and have also been found useful in training offline models. Opcodes are machine code instructions carried out by the CPU, such as performing arithmetic operations and accessing memory locations.
\end{itemize}


And the following dynamic features:
\begin{itemize}
    \item \textbf{System calls:} The most commonly used dynamic feature in the literature. System calls are requests made by the application to the operating system's kernel to perform tasks during the execution of the application. 
    \item \textbf{API calls:} Dynamically obtained API calls are API calls made by the application during run-time. A prior list of API calls to be tracked is provided to the analyzer, which then tracks them during the execution of an application.
\end{itemize}

Finally, for our hybrid feature set, we use system calls, permissions, and opcodes. Static API calls were excluded from our hybrid feature sets since they undergo the most frequent changes, with additions and removals in almost every update. Furthermore, in previous work, we found there was no significant benefit to using these alongside dynamic features.

\subsection{Evaluation}

Evaluation of OL models differs from techniques used to evaluate traditional offline ML models. OL models are trained as new data arrives, and are therefore also evaluated incrementally. There are two common approaches to doing this. The first is \textit{holdout}: training sets are used to incrementally train the classifier and holdout sets are then used periodically to test the model. However, this presents challenges in terms of selecting the best holdout set interval and determining how many holdout sets to use during the evaluation process. The other approach is \textit{progressive}: each individual data item, in our case Android application, is used first to test the existing model and then to incrementally train the model. This allows the performance metrics to be updated incrementally, and provides a clearer picture regarding how the model will perform in a real world scenario over time.

In this work, we use progressive evaluation, and incrementally measure the accuracy, F1-score, precision, and the confusion matrix of each OL model. However, the progressive evaluation process is slightly different depending upon whether we are evaluating a model within a progressive validation context, i.e. where it is assumed that a correct label is instantly available, or within the more realistic delayed progressive validation context. In the latter, each new data item is still used to test the model; however, it is not used to incrementally train the model until a label becomes available. For a benign application, the release date is used to determine when it is used to test the model, and the date of its VirusTotal report is used as the date when it becomes labelled and is used to train the model. For malware, the release date is not normally available, and so we estimate it based on the average period it takes for a newly released application to be assigned a VirusTotal report. The average time it took for applications in our dataset to be assigned a VirusTotal report was 40 days, and hence, for malware, the release date is estimated as 40 days before the date on its VirusTotal report.

An initial seed dataset is used to train the models before incremental learning begins. That is, all applications with a release date prior to the date that incremental learning begins are used to train the initial model. This seed dataset comprises 4,199 malware and 3,925 benign applications. We also present the results of the models trained on just the initial dataset and not updating them with new data in Appendix \ref{appendix}.

\subsection{Active learning Framework}

\begin{figure}[t]
\includegraphics[width=\columnwidth]{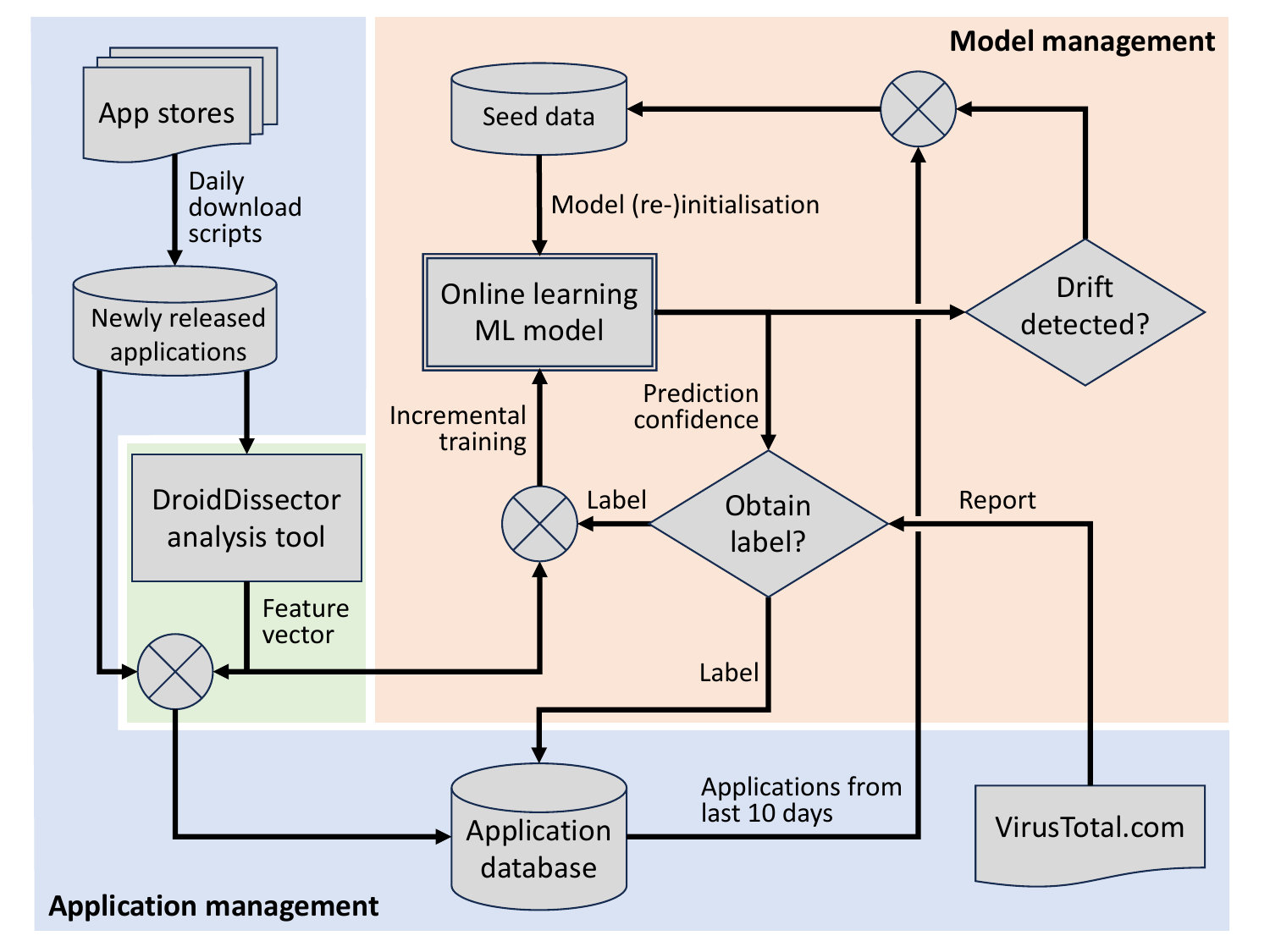}
\centering
\caption{Our active learning framework}
\label{fig:olframework}
\end{figure}

In this section, we introduce our novel semi-supervised active learning framework. The concept behind active learning is that a model should not train on all data samples, but should actively select which data samples it learns from in order to improve its performance. The idea is that, by focusing on the most informative data samples, it can improve its learning rate and accuracy, whilst also requiring less labelled data. In our framework, we implement this idea by incrementally training a model only on newly released Android applications for which the model currently has a low confidence in its prediction, specifically when the model's confidence is below a threshold of 80\%. 

The model's confidence in a prediction is determined by using the library's ``predict\_proba" method. In PA, the confidence of prediction is measured by the degree of separation between the predicted class and other classes. Typically in Hoeffding trees, confidence is calculated by the number of training instances in the leaf node of the predicted class. Confidence in adaptive RF is determined by the fraction of trees in the ensemble that agrees with the prediction. KNN uses a voting scheme where each nearest neighbour votes for the prediction and the fraction of votes is used for the confidence of the prediction. The probability from Gaussian Naïve Bayes is used as the confidence of its prediction.   

The framework architecture is depicted in Figure \ref{fig:olframework}. The framework comprises four modules: application management, automated analysis, model management, and a database. The framework was designed with practical deployment in mind, and uses tools we have previously built, including dataset collection tools and static and dynamic analysis tools. The four modules are described below.

\paragraph{Application management} This module is responsible for finding and downloading Android applications. It uses several third-party application crawler scripts to download newly released applications, and runs these scripts as a cron job every day, returning applications that were released in the past 24 hours. It is also responsible for calling the VirusTotal API to obtain reports, when required.

\paragraph{Automated analysis} We use our existing tool DroidDissector \citep{muzaffar2023droiddissector} to carry out static and/or dynamic analysis of each downloaded application. The tool extracts all relevant features for a given application, and this module then returns a feature vector which can be used in the testing or training of models.

\paragraph{Model management} This module maintains the OL models, and is responsible for training and evaluation, using the feature vector obtained from the automated analysis module. The models are built using river \citep{riverOl}, a python OL library. The pickle library is used to store models between sessions.

\paragraph{Database} The database stores downloaded applications; their label, if a report has been obtained from VirusTotal; release date, actual or estimated; MD5, SHA-1 and SHA-256 hashes; and their feature vector. The latter avoids repeated use of the computationally-expensive automated analysis tools when the same application is seen multiple times.

\section{Results}

We begin by presenting results from training standard OL models using the two evaluation scenarios: progressive validation in section \ref{sec:progressive} and delayed progressive validation in section \ref{sec:delayed}. These then provide a baseline against which to compare the results of our active learning framework, which are presented in section \ref{sec:active}. In Appendix \ref{appendix}, we also give the outcomes of the models that were just trained on the initial dataset, without being updated with fresh data. The findings strongly emphasize the importance of updating the OL models to sustain their performance over time.

\subsection{Progressive Validation} \label{sec:progressive}

\begin{figure}
     \centering
     \begin{subfigure}{\textwidth}
     \centering
         \includegraphics[width=\columnwidth]{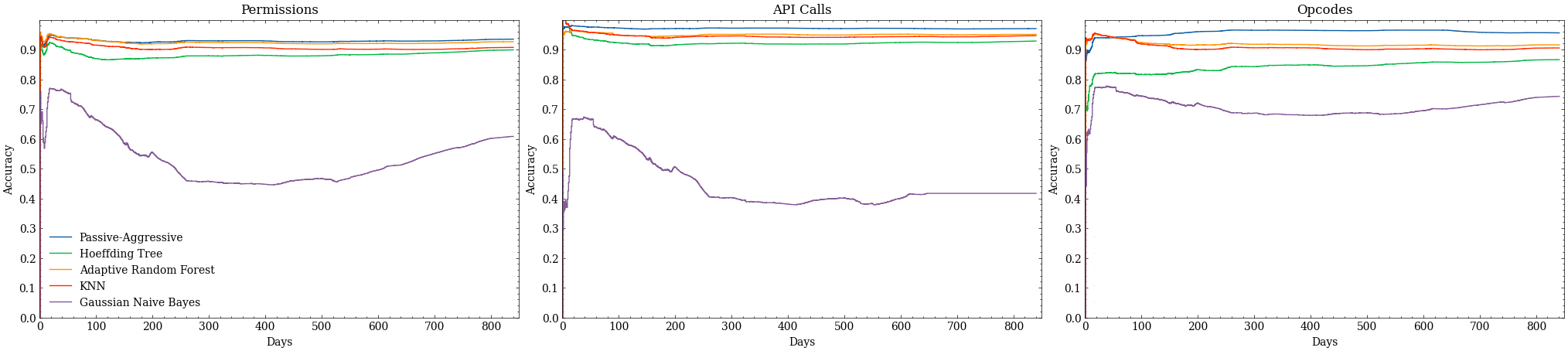}
         \caption{Accuracy}
         \label{fig:psa}
     \end{subfigure}
    
     \begin{subfigure}{\textwidth}
         \centering
         \includegraphics[width=\columnwidth]{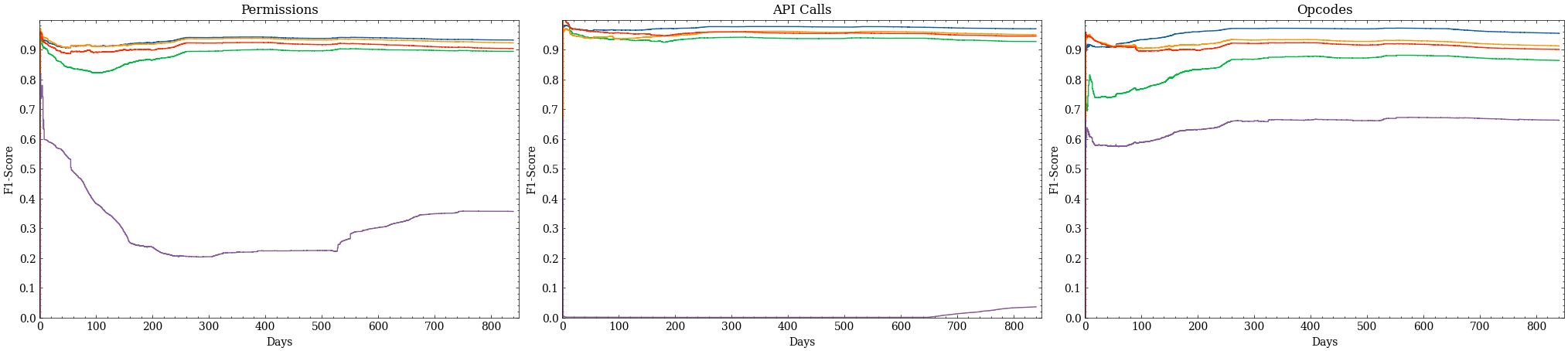}
         \caption{F1-Score}
         \label{fig:psf}
     \end{subfigure}
 
     \begin{subfigure}{\textwidth}
         \centering
         \includegraphics[width=\columnwidth]{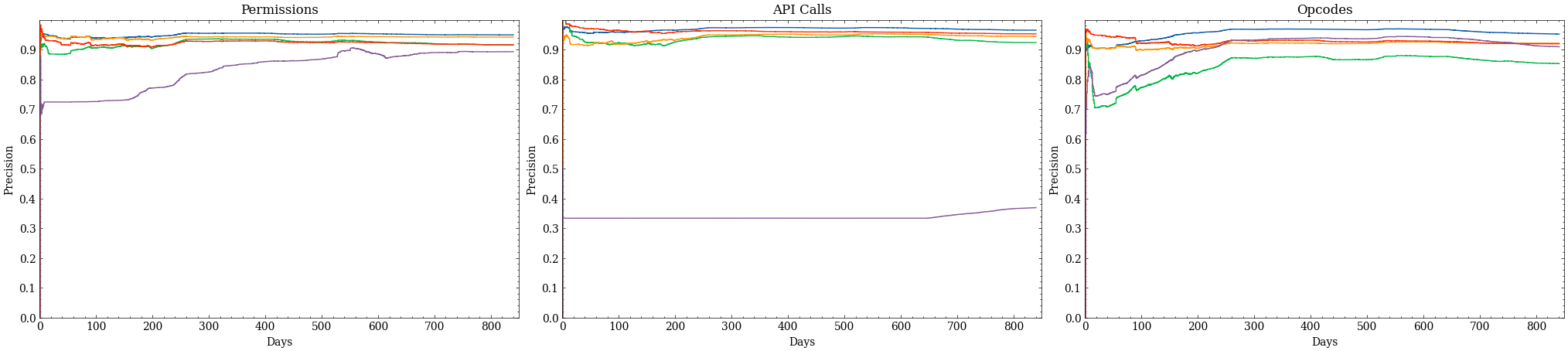}
         \caption{Precision}
         \label{fig:psp}
     \end{subfigure}

     \begin{subfigure}{\textwidth}
         \centering
         \includegraphics[width=\columnwidth]{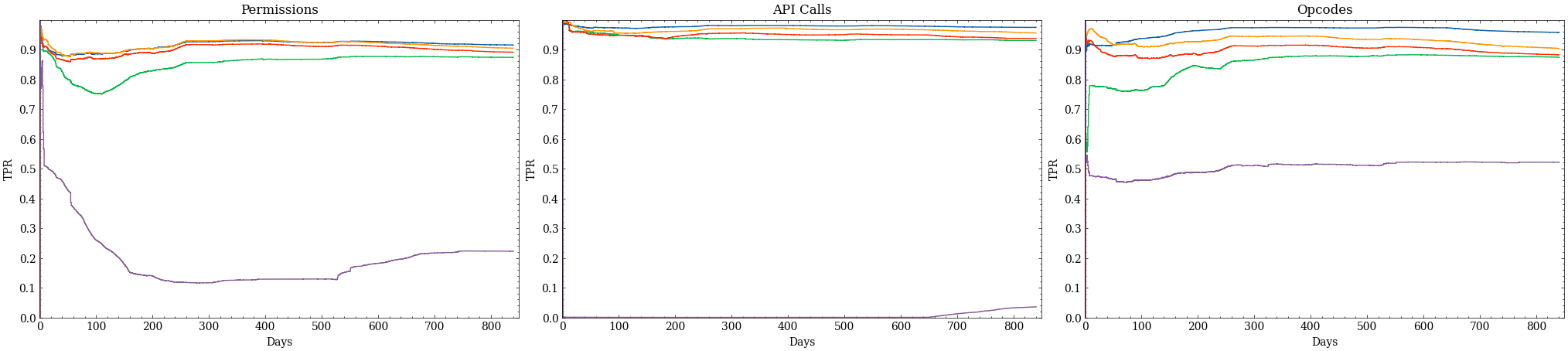}
         \caption{TPR}
         \label{fig:pst}
     \end{subfigure}
        \caption{Progressive validation results on OL models trained on static features}
        \label{fig:progressive}
\end{figure}

\begin{figure}
     \centering
     \begin{subfigure}{\textwidth}
     \centering
         \includegraphics[width=\columnwidth]{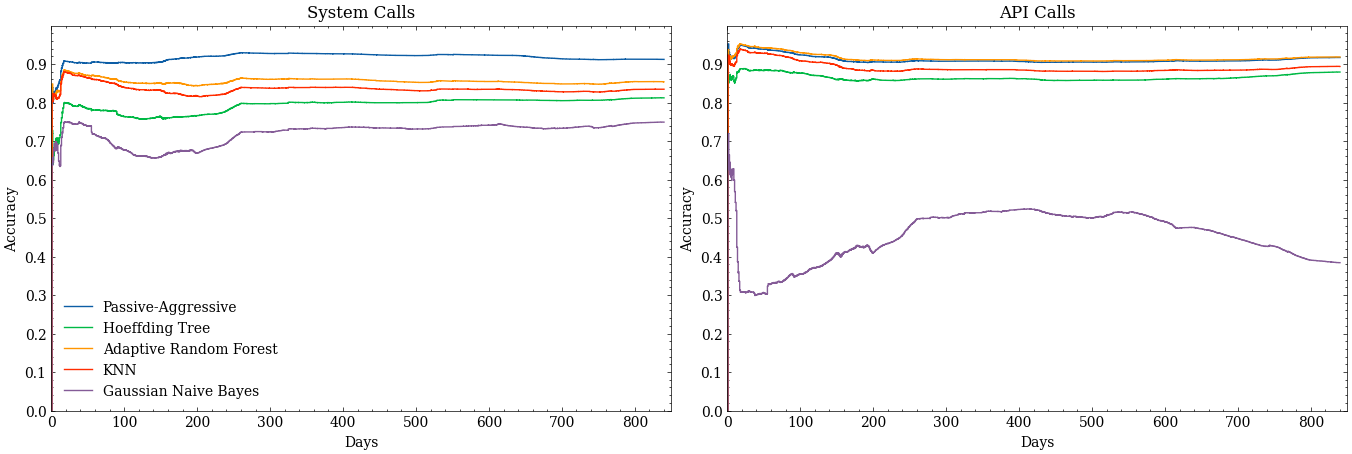}
         \caption{Accuracy}
         \label{fig:pda}
     \end{subfigure}
    
     \begin{subfigure}{\textwidth}
         \centering
         \includegraphics[width=\columnwidth]{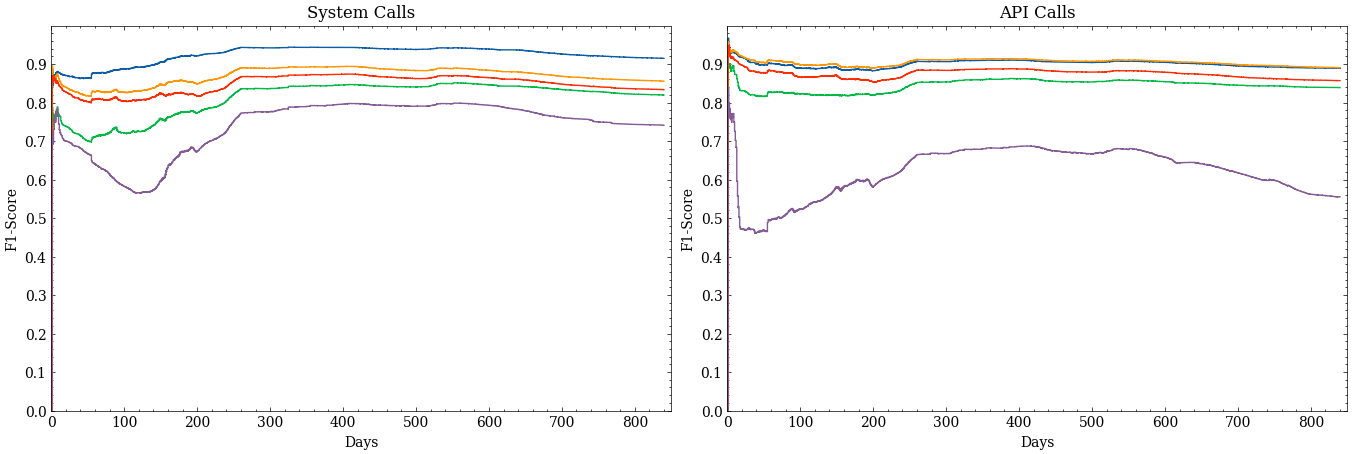}
         \caption{F1-Score}
         \label{fig:pdf}
     \end{subfigure}
 
     \begin{subfigure}{\textwidth}
         \centering
         \includegraphics[width=\columnwidth]{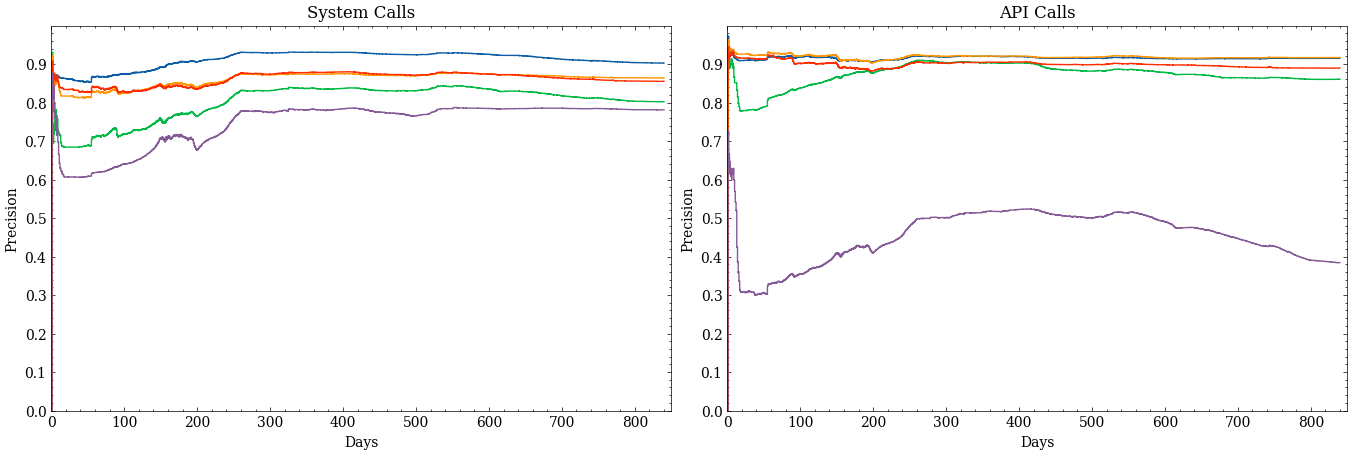}
         \caption{Precision}
         \label{fig:pdp}
     \end{subfigure}

     \begin{subfigure}{\textwidth}
         \centering
         \includegraphics[width=\columnwidth]{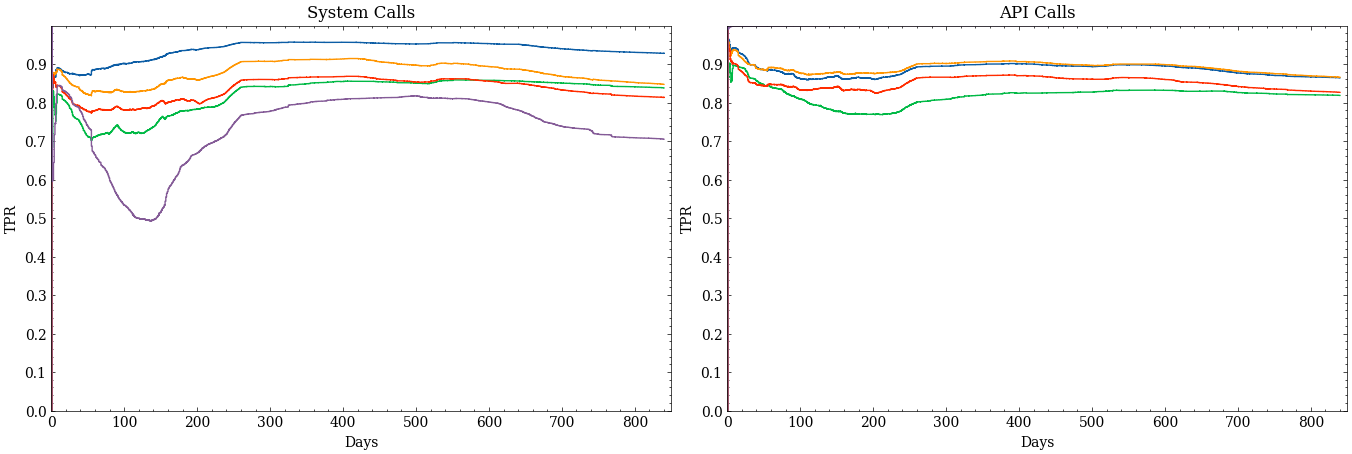}
         \caption{TPR}
         \label{fig:pdt}
     \end{subfigure}

        \caption{Progressive validation results on OL models trained on dynamic features}
        \label{fig:progressivedynamic}
\end{figure}

\begin{figure}
     \centering
       \begin{subfigure}{\textwidth}
     \centering
         \includegraphics[width=\columnwidth]{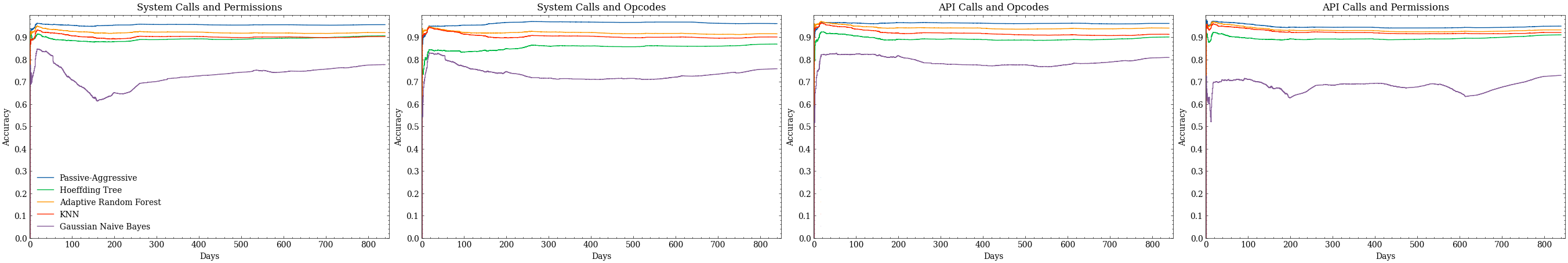}
         \caption{Accuracy}
         \label{fig:pha}
     \end{subfigure}
    
     \begin{subfigure}{\textwidth}
         \centering
         \includegraphics[width=\columnwidth]{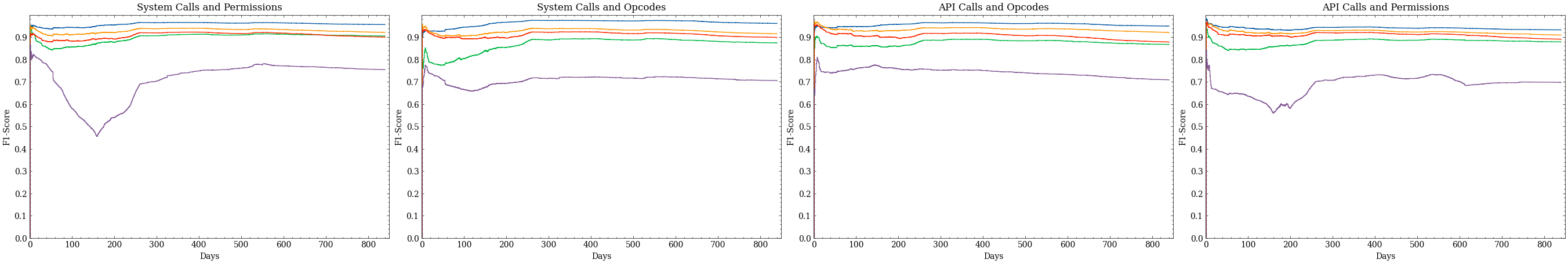}
         \caption{F1-Score}
         \label{fig:phf}
     \end{subfigure}
 
     \begin{subfigure}{\textwidth}
         \centering
         \includegraphics[width=\columnwidth]{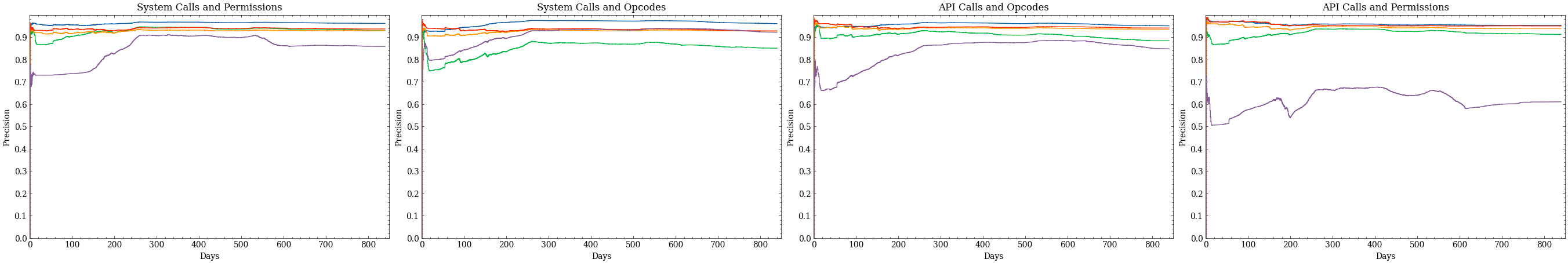}
         \caption{Precision}
         \label{fig:php}
     \end{subfigure}

     \begin{subfigure}{\textwidth}
         \centering
         \includegraphics[width=\columnwidth]{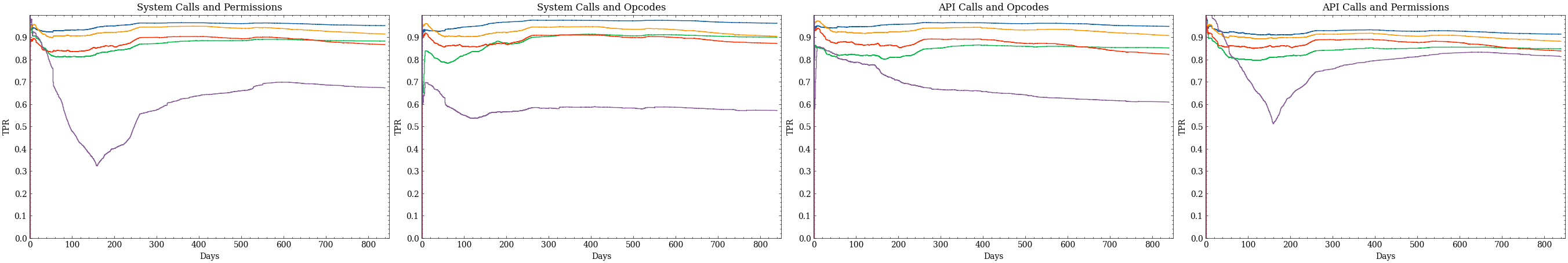}
         \caption{TPR}
         \label{fig:pht}
     \end{subfigure}

        \caption{Progressive validation results on OL models trained on hybrid features}
        \label{fig:progressivehybrid}
\end{figure}

We first evaluated standard OL models using the ideal progressive validation scenario, i.e. when each new data item is assumed to immediately have a label, and can therefore be used for both incrementally training and testing each model. Table \ref{tab:p_table} summarises the results, showing the average metrics for the best performing class of OL model when different feature sets are used. Figures \ref{fig:progressive}--\ref{fig:progressivehybrid} plot detailed results for the 5 types of OL model when static, dynamic and hybrid feature sets are used, highlighting how the 4 evaluation metrics vary over the course of the online learning period.

\begin{table}[t]
    \caption{Summary of performance under the progressive validation scenario. For each feature set, the mean and standard deviations of metrics of the best-performing OL model are shown.  Horizontal rules separate the static, dynamic and hybrid feature sets.}
    \label{tab:p_table}
    \centering
    \resizebox{\columnwidth}{!}{%
        \begin{tabular}{@{}llllllll@{}}
            \toprule
            \textbf{Feature}             & \textbf{OL Model} & \textbf{Accuracy}  & \textbf{F1-Score} & \textbf{Precision}& \textbf{TPR} & \textbf{Drifts} \\
            \midrule
            Permissions                  & PA                & 0.928$ \pm 0.016$      & 0.932$ \pm 0.015$ & 0.948$ \pm 0.018$       & 0.916$ \pm 0.017$      & 16                                 \\
            API Calls                       & PA & \textbf{0.971}$ \pm 0.007$ & \textbf{0.973}$ \pm 0.008$ & \textbf{0.963}$ \pm 0.008$ & \textbf{0.978}$ \pm 0.007$ & 14   \\
            Opcodes                      & PA                & 0.952$ \pm 0.029$     & 0.955$ \pm 0.031$ & 0.952$ \pm 0.030$        & 0.958$ \pm 0.031$                               & 13                  \\       
            \midrule
            System Calls                 & PA                & 0.911$ \pm 0.031$       & 0.919$ \pm 0.037$ & 0.901$ \pm 0.032$      & 0.932$ \pm 0.042$                 & 17         \\
            Dynamic API Calls                & PA                & 0.912$ \pm 0.013$   & 0.900$ \pm 0.016$ & 0.915$ \pm 0.010$          & 0.889$ \pm 0.025$                & 18     \\    
            \midrule
            System Calls and Permissions & PA                & 0.952$ \pm 0.019$     & 0.957$ \pm 0.022$ & 0.961$ \pm 0.019$        & 0.952$ \pm 0.025$            & 14   \\
            System Calls and Opcodes     & PA                & 0.958$ \pm 0.027$   & 0.951$ \pm 0.028$ & 0.961$ \pm 0.028$          & 0.952$ \pm 0.028$            & 14    \\  
                        Dynamic API Calls and Opcodes     & PA                & 0.959$ \pm 0.025$   & 0.954$ \pm 0.024$ & 0.955$ \pm 0.024$          & 0.954$ \pm 0.025$              & 15   \\  
            Dynamic API Calls and Permissions & PA                & 0.947$ \pm 0.013$     & 0.940$ \pm 0.013$ & 0.957$ \pm 0.013$        & 0.924$ \pm 0.015$             & 14   \\

            \bottomrule
        \end{tabular}%
    }
\end{table}

The best models led to a mean accuracy of about 97\% across the online learning period. In all cases, PA appears to be the most effective model, although adaptive RFs and KNN also perform competitively in most cases. Naive Bayes models generally perform poorly, suggesting that feature interactions are important. In terms of feature sets, static API calls lead to the most accurate models, which reflects similar understanding from the training of offline models \citep{Muzaffar2023}. The static API call models also have the lowest variance, and this can be seen in the plots as a relatively consistent level of accuracy throughout the online learning period.

However, it is expensive to build models from static API feature sets, since the dimensionalities of these features sets are very large --- 134,207 in our case. During our experiments, we observed a number of frozen runs due to the machine running low on resources. In this respect, it is also worth considering other, cheaper, feature sets. Permissions are very easy to extract, but performance was significantly lower with permissions-based models. Opcode feature sets seem to be a better option, since they lead to relatively good accuracies of around 95\%, whilst being a lot lower in dimensionality --- there were 218 opcodes in our feature set.

Dynamic feature sets do not appear to lead to good models, at least when used individually. Given that they are a lot more expensive to extract than static features, this seems to indicate against building models from dynamic features alone. The hybrid feature sets all lead to models with a similar level of performance, and interestingly combining system calls with permissions leads to models that are significantly better than those trained using either feature alone. Given that both of these features have low dimensionality, this combination could be practically useful, at least in terms of training effort if not feature extraction cost.
 
\subsection{Delayed Progressive Validation} \label{sec:delayed}

\begin{figure}
     \centering
          \begin{subfigure}{\textwidth}
     \centering
         \includegraphics[width=\columnwidth]{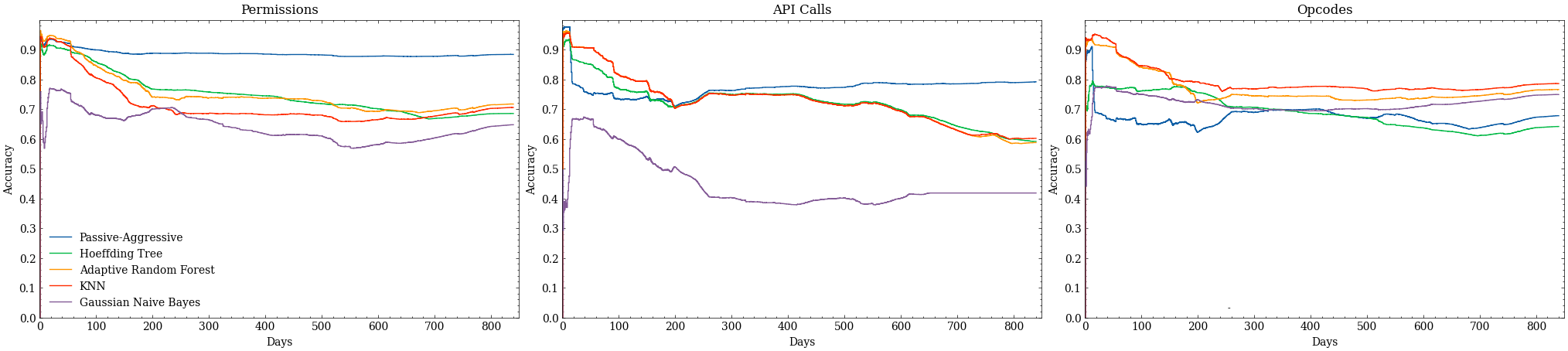}
         \caption{Accuracy}
         \label{fig:dsa}
     \end{subfigure}
    
     \begin{subfigure}{\textwidth}
         \centering
         \includegraphics[width=\columnwidth]{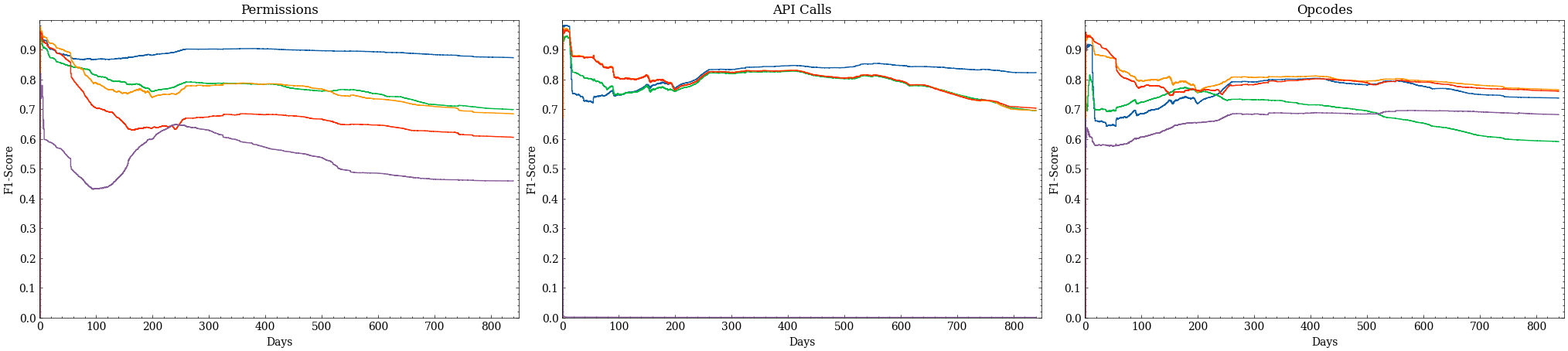}
         \caption{F1-Score}
         \label{fig:dsf}
     \end{subfigure}
 
     \begin{subfigure}{\textwidth}
         \centering
         \includegraphics[width=\columnwidth]{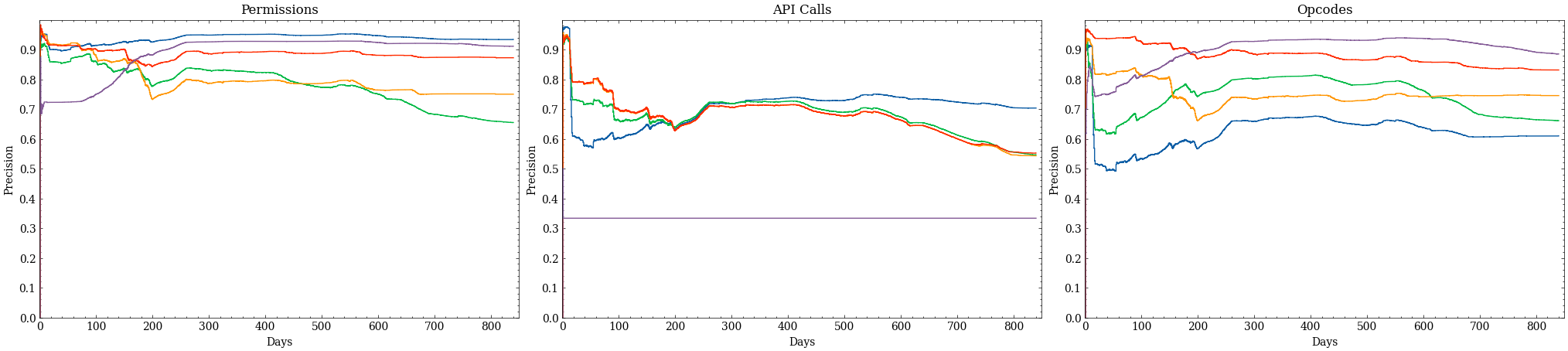}
         \caption{Precision}
         \label{fig:dsp}
     \end{subfigure}

     \begin{subfigure}{\textwidth}
         \centering
         \includegraphics[width=\columnwidth]{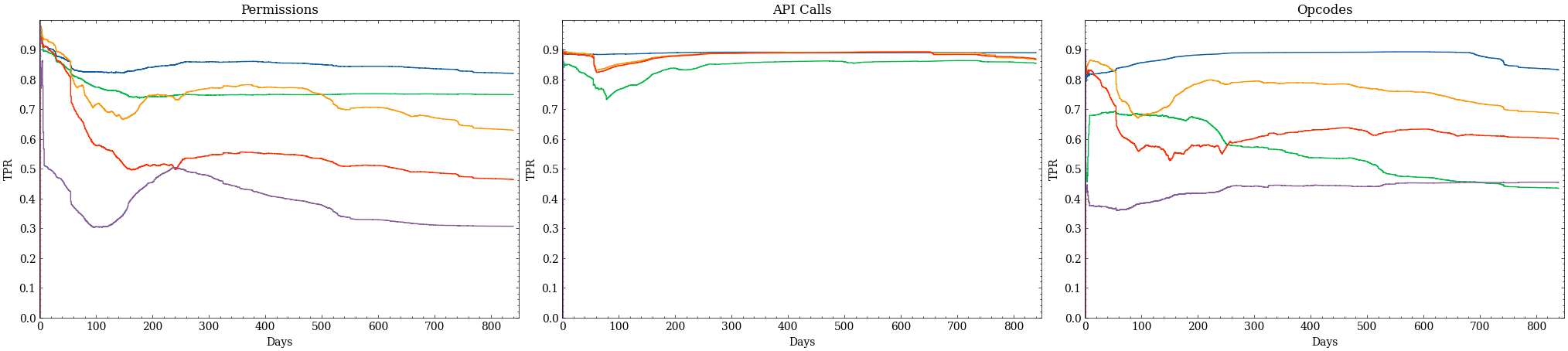}
         \caption{TPR}
         \label{fig:dst}
     \end{subfigure}
        \caption{Delayed progressive validation results on OL models trained on static features}
        \label{fig:delayed}
\end{figure}

\begin{figure}
     \centering
       \begin{subfigure}{\textwidth}
     \centering
         \includegraphics[width=\columnwidth]{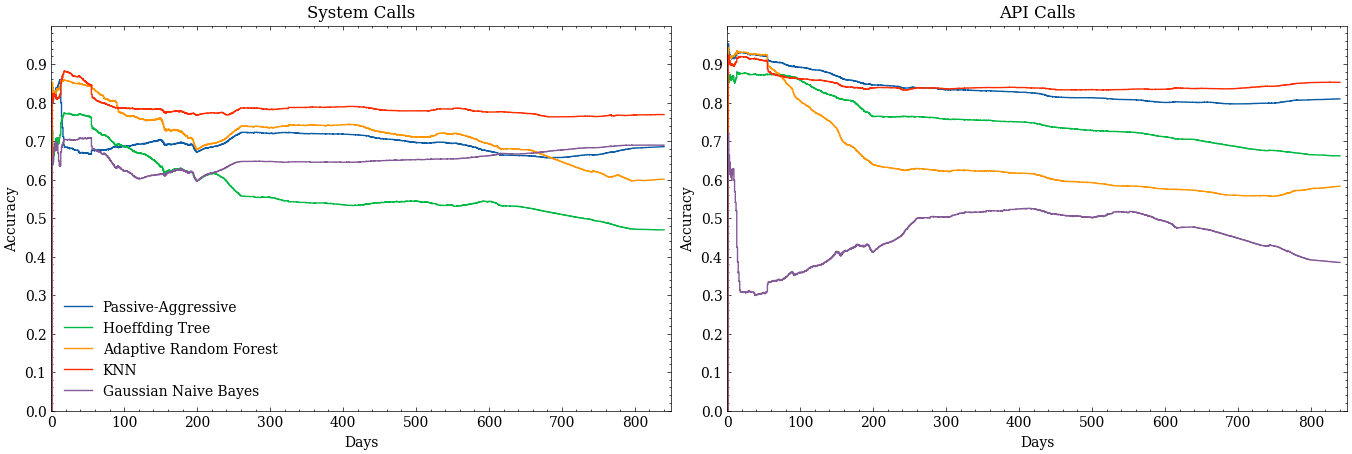}
         \caption{Accuracy}
         \label{fig:dda}
     \end{subfigure}
    
     \begin{subfigure}{\textwidth}
         \centering
         \includegraphics[width=\columnwidth]{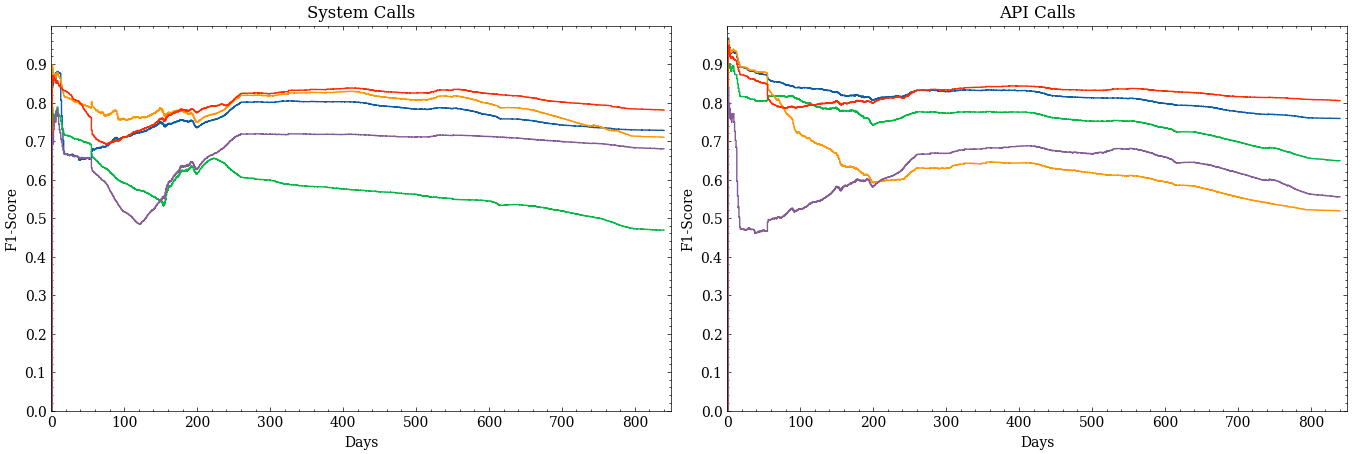}
         \caption{F1-Score}
         \label{fig:ddf}
     \end{subfigure}
 
     \begin{subfigure}{\textwidth}
         \centering
         \includegraphics[width=\columnwidth]{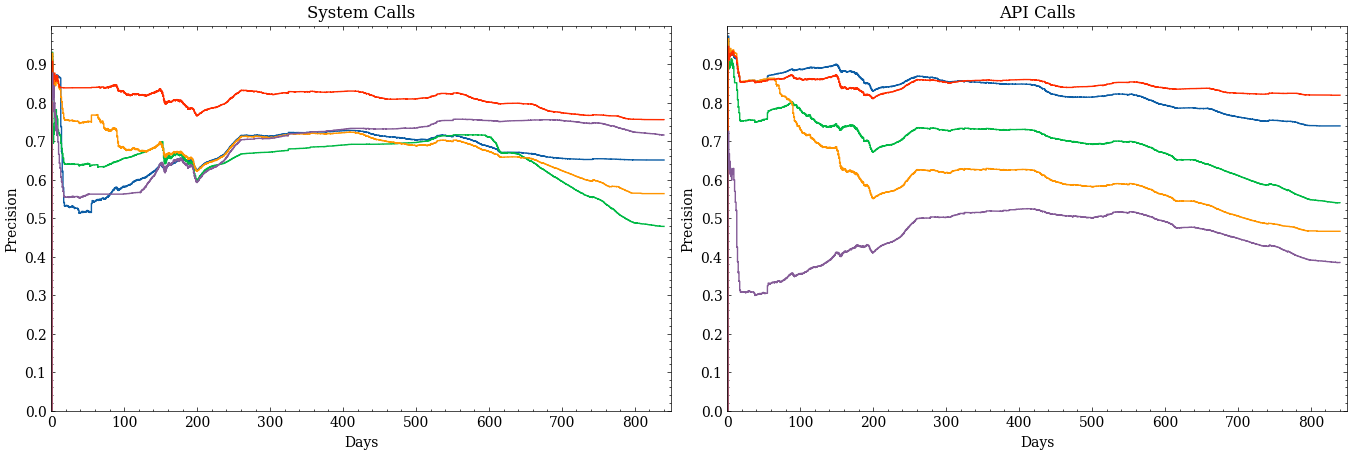}
         \caption{Precision}
         \label{fig:ddp}
\end{subfigure}

     \begin{subfigure}{\textwidth}
         \centering
         \includegraphics[width=\columnwidth]{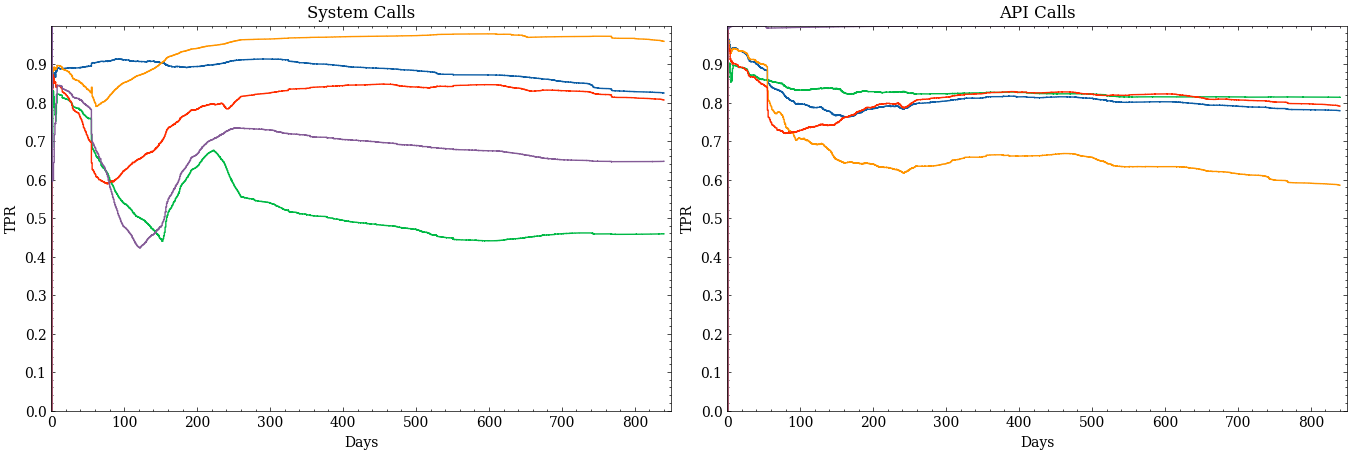}
         \caption{TPR}
         \label{fig:ddt}
     \end{subfigure}

        \caption{Delayed progressive validation results on OL models trained on dynamic features}
        \label{fig:delayeddynamic}
\end{figure}

\begin{figure}
     \centering
    \begin{subfigure}{\textwidth}
     \centering
         \includegraphics[width=\columnwidth]{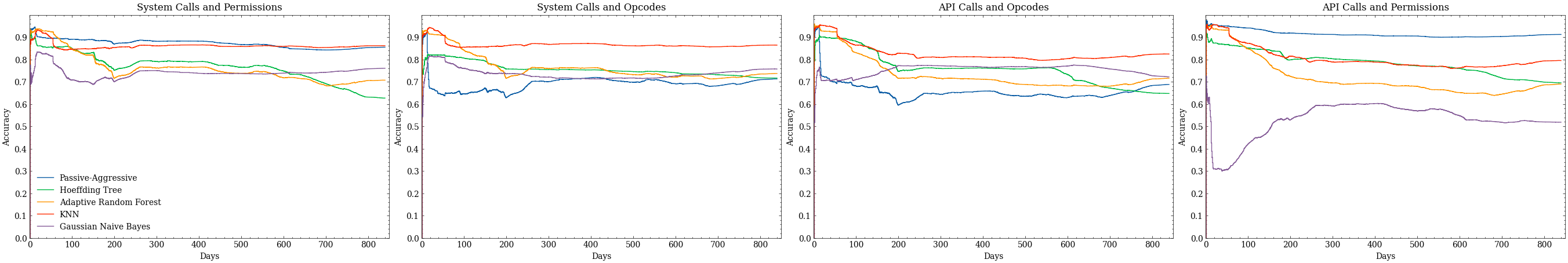}
         \caption{Accuracy}
         \label{fig:dha}
     \end{subfigure}
    
     \begin{subfigure}{\textwidth}
         \centering
         \includegraphics[width=\columnwidth]{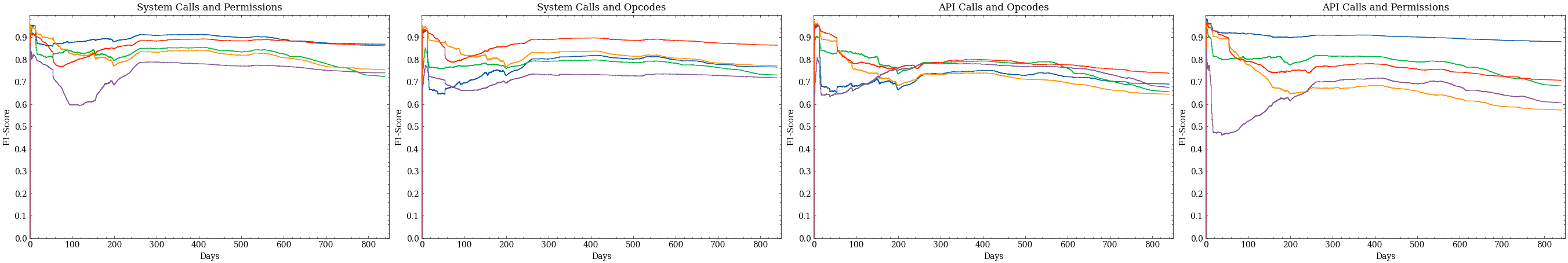}
         \caption{F1-Score}
         \label{fig:dhf}
     \end{subfigure}
 
     \begin{subfigure}{\textwidth}
         \centering
         \includegraphics[width=\columnwidth]{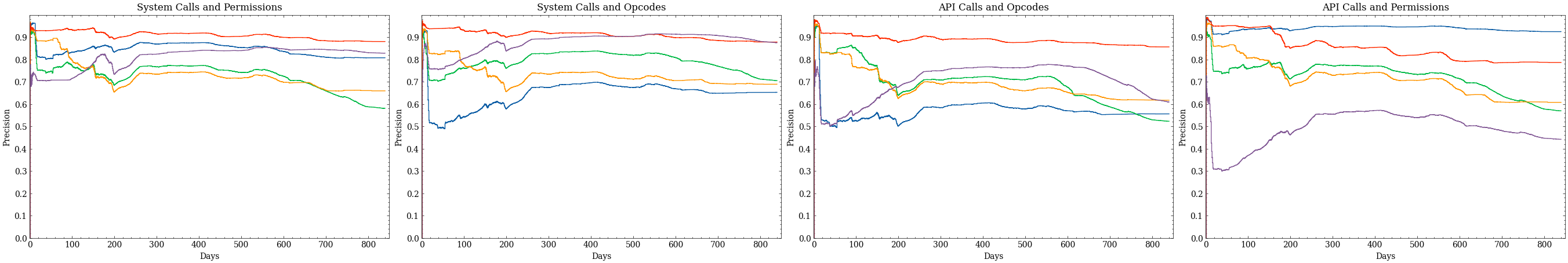}
         \caption{Precision}
         \label{fig:dhp}
     \end{subfigure}

     \begin{subfigure}{\textwidth}
         \centering
         \includegraphics[width=\columnwidth]{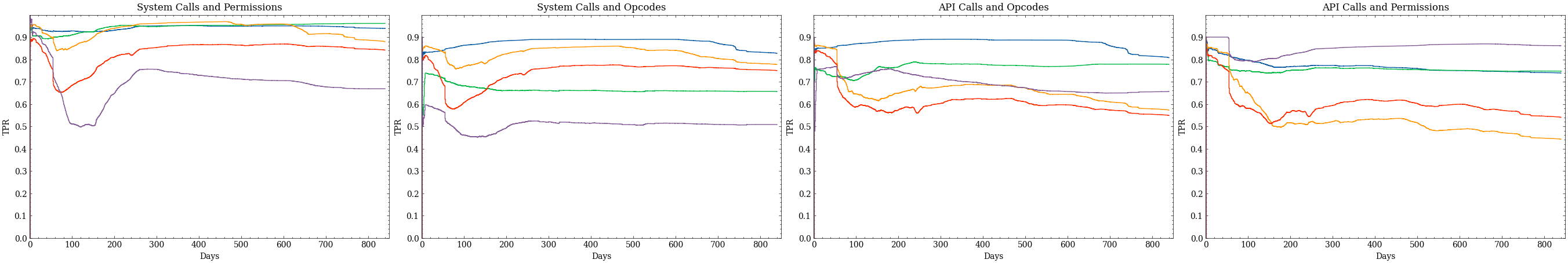}
         \caption{TPR}
         \label{fig:dht}
     \end{subfigure}

        \caption{Delayed progressive validation results on OL models trained on hybrid features}
        \label{fig:delayedhybrid}
\end{figure}

We next consider the more realistic scenario in which labels are only available a period of time after the date at which an application is released. This delays the time between an application becoming available to users and the application being available to incrementally train a malware detection model. Table \ref{tab:d_p_table} summarises the results, showing the average metrics for the best performing class of OL model when different feature sets are used. Figures \ref{fig:delayed}--\ref{fig:delayedhybrid} plot detailed results for the 5 types of OL model when static, dynamic and hybrid feature sets are used.

\begin{table}[t]
    \caption{Summary of performance under the delayed progressive validation scenario. For each feature set, the mean and standard deviations of metrics of the best-performing OL model are shown. Horizontal rules separate the static, dynamic and hybrid feature sets.}
    \label{tab:d_p_table}
    \centering
    \resizebox{\columnwidth}{!}{%
        \begin{tabular}{@{}llllllll@{}}
            \toprule
            \textbf{Feature}             & \textbf{OL Model} & \textbf{Accuracy}  & \textbf{F1-Score} & \textbf{Precision}& \textbf{TPR} & \textbf{Drifts} \\
            \midrule
            Permissions                  &  PA                & 0.888$ \pm 0.040$      & 0.890$ \pm 0.039$  & \textbf{0.937}$ \pm 0.040$          & 0.848$ \pm 0.038$                        & 39               \\
            API Calls  & KNN & 0.783$ \pm 0.015$   & 0.830$ \pm 0.069$  & 0.719$ \pm 0.090$  & 0.901$ \pm 0.024$    & 45   
            \\
            Opcodes                      & KNN               & 0.839$ \pm 0.047$     & 0.840$ \pm 0.050$  &  0.892$ \pm 0.042$       & 0.879$ \pm 0.065$                        & 34            \\
            \midrule
            System Calls                  & KNN               & 0.782$ \pm 0.034$      & 0.801$ \pm 0.037$ &  0.802$ \pm 0.038$     & 0.803$ \pm 0.063$                       & 52                        \\
            
            Dynamic API Calls                &  KNN               & 0.832$ \pm 0.037$  & 0.818$ \pm 0.044$ & 0.834$ \pm 0.038$            & 0.802$ \pm 0.059$                  & 44                        \\
            \midrule
            System Calls and Permissions &  PA               & 0.872$ \pm 0.036$    & 0.891$ \pm 0.031$ & 0.847$ \pm 0.042$         & \textbf{0.941}$ \pm 0.026$                        & 34             \\
            System Calls and Opcodes    & KNN               & 0.853$ \pm 0.057$       & 0.860$ \pm 0.053$ & 0.897$ \pm 0.080$     & 0.825$ \pm 0.033$      & 33        \\     

            Dynamic API Calls and Opcodes     & KNN                & 0.832$ \pm 0.077$   & 0.790$ \pm 0.064$ & 0.889$ \pm 0.098$          & 0.644$ \pm 0.034$           & 32     \\  
                        Dynamic API Calls and Permissions & PA                & \textbf{0.908}$ \pm 0.032$     & \textbf{0.902}$ \pm 0.034$ & 0.912$ \pm 0.041$        & 0.770$ \pm 0.052$          & 36      \\
            \bottomrule
        \end{tabular}%
    }
\end{table}

It is clear that the results are very different to those presented in the previous section. Static API calls are no longer the best feature --- in fact, they are one of the worst --- and Fig. \ref{fig:dsa} shows that this is due to a rapid degradation in performance after the initially good performance on the seed data. All the models show this degradation to a certain extent, suggesting there is significant concept drift during the period between application release and application labelling.

However, it is interesting to see that some features are more robust to this process than others. Permissions, for example, seem to be relatively robust. An explanation for this is that the usage of permissions stays fairly constant over time, since there are few of these and they are concerned with the high-level behaviour of an application. The use of API calls, on the other hand, could change quite rapidly as attackers respond to anti-malware defence strategies and as APIs are added and removed from Android SDK overtime, more frequently than permissions. 

Permissions also lead to the models with the highest precision; that is, if the model says an application is malware, then it probably is. When combined with system calls, the resulting models have the highest TPR, but a lower precision; that is, the addition of system calls makes it more likely that malware will be correctly identified as such, but the number of false positives will also be higher.

In terms of model accuracy, the most effective approach appears to be combining the permissions feature set with the dynamic API calls feature set. This combination also showed the least degradation in accuracy during the online learning period. However, the accuracy of the best models is significantly lower than in the progressive validation scenario, with a loss in mean accuracy of around 6\%. PA is still the best performing model overall, though KNN now performs better with the majority of the feature sets. Overall it seems that when OL models are reevaluated in a more realistic scenario, the outcomes are quite different, and observations about features and models often do not generalise between the two.





\subsection{Active learning Framework} \label{sec:active}

\begin{figure}
     \centering
         \begin{subfigure}{\textwidth}
     \centering
         \includegraphics[width=\columnwidth]{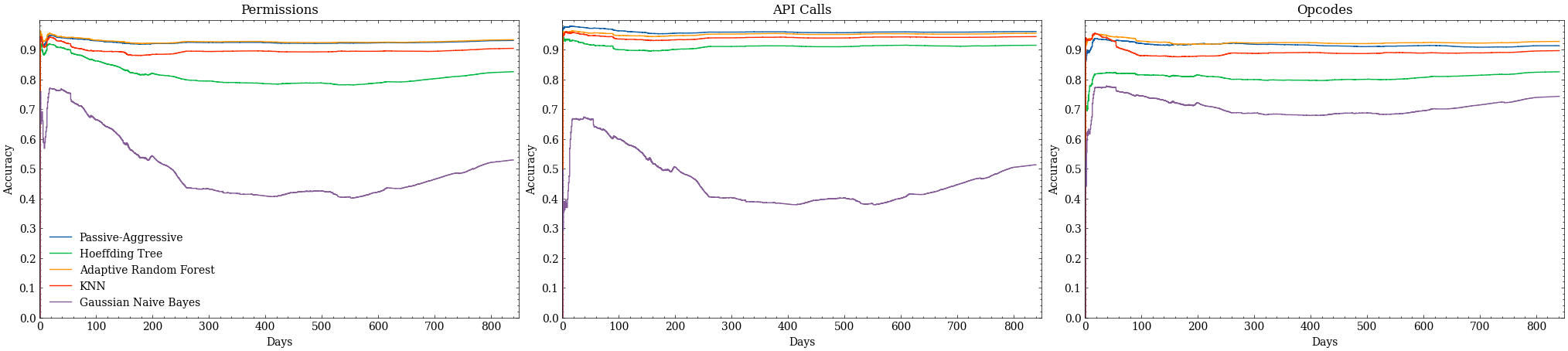}
         \caption{Accuracy}
         \label{fig:asa}
     \end{subfigure}
    
     \begin{subfigure}{\textwidth}
         \centering
         \includegraphics[width=\columnwidth]{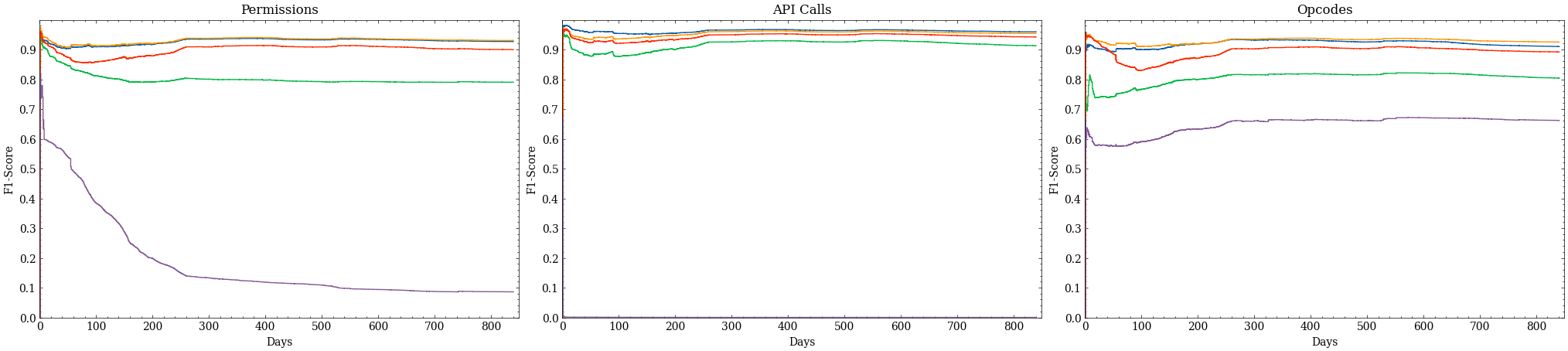}
         \caption{F1-Score}
         \label{fig:asf}
     \end{subfigure}
 
     \begin{subfigure}{\textwidth}
         \centering
         \includegraphics[width=\columnwidth]{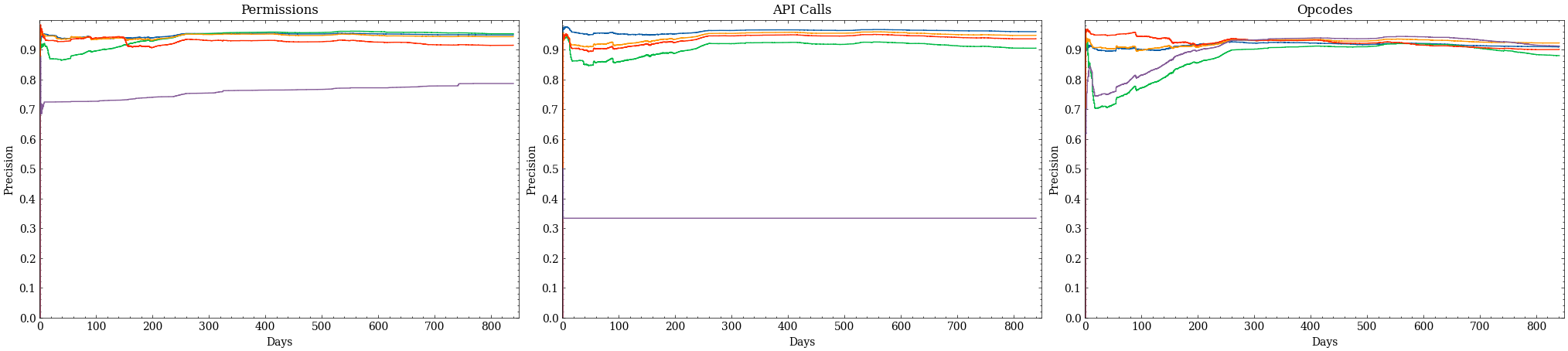}
         \caption{Precision}
         \label{fig:asp}
     \end{subfigure}

     \begin{subfigure}{\textwidth}
         \centering
         \includegraphics[width=\columnwidth]{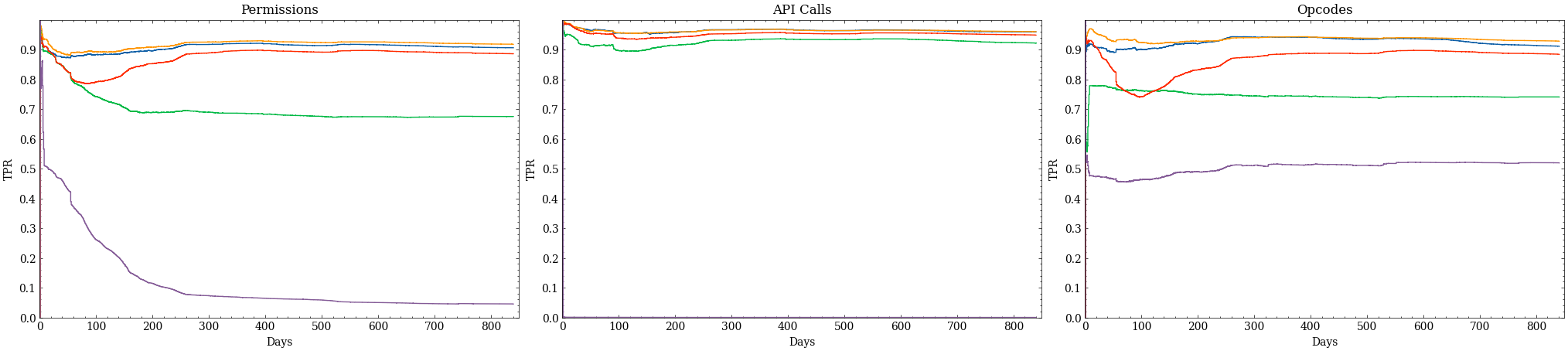}
         \caption{TPR}
         \label{fig:ast}
     \end{subfigure}

        \caption{Active learning framework results on OL models trained on static features}
        \label{fig:activestatic}
\end{figure}

\begin{figure}
     \centering
       \begin{subfigure}{\textwidth}
     \centering
         \includegraphics[width=\columnwidth]{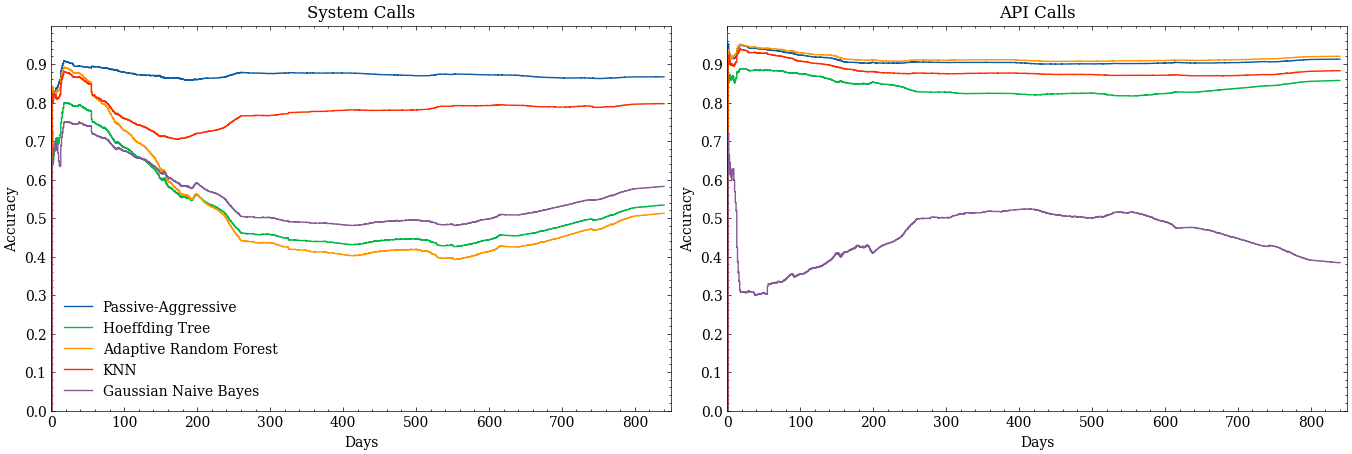}
         \caption{Accuracy}
         \label{fig:ada}
     \end{subfigure}
    
     \begin{subfigure}{\textwidth}
         \centering
         \includegraphics[width=\columnwidth]{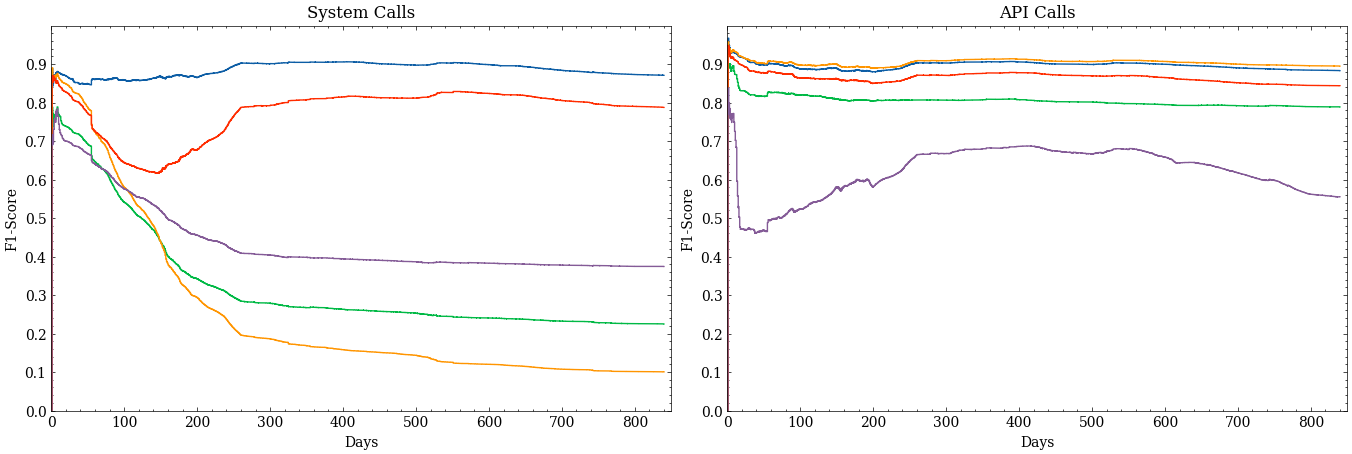}
         \caption{F1-Score}
         \label{fig:adf}
     \end{subfigure}
 
     \begin{subfigure}{\textwidth}
         \centering
         \includegraphics[width=\columnwidth]{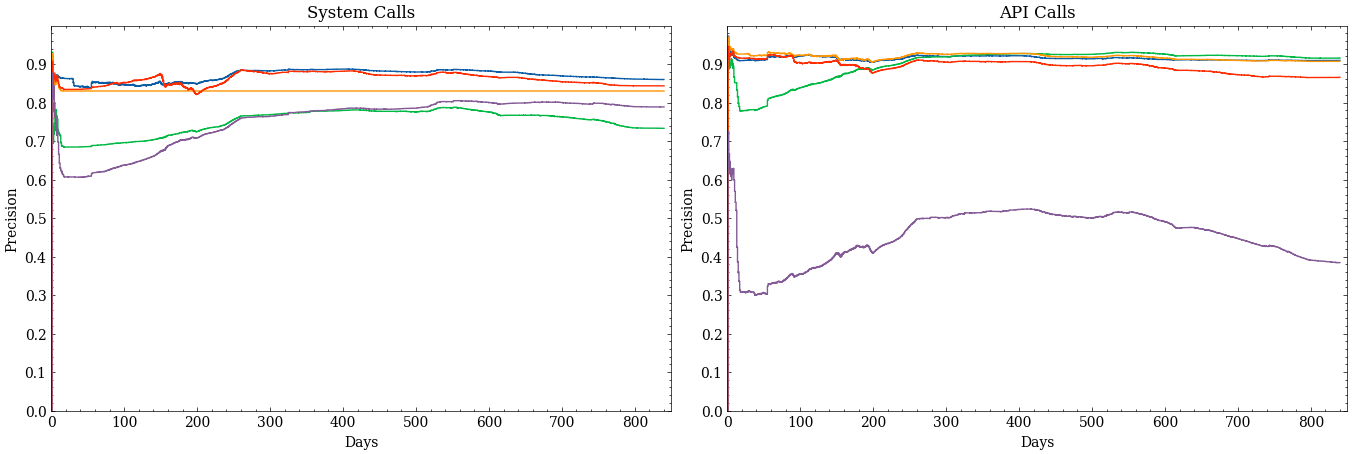}
         \caption{Precision}
         \label{fig:adp}
     \end{subfigure}

     \begin{subfigure}{\textwidth}
         \centering
         \includegraphics[width=\columnwidth]{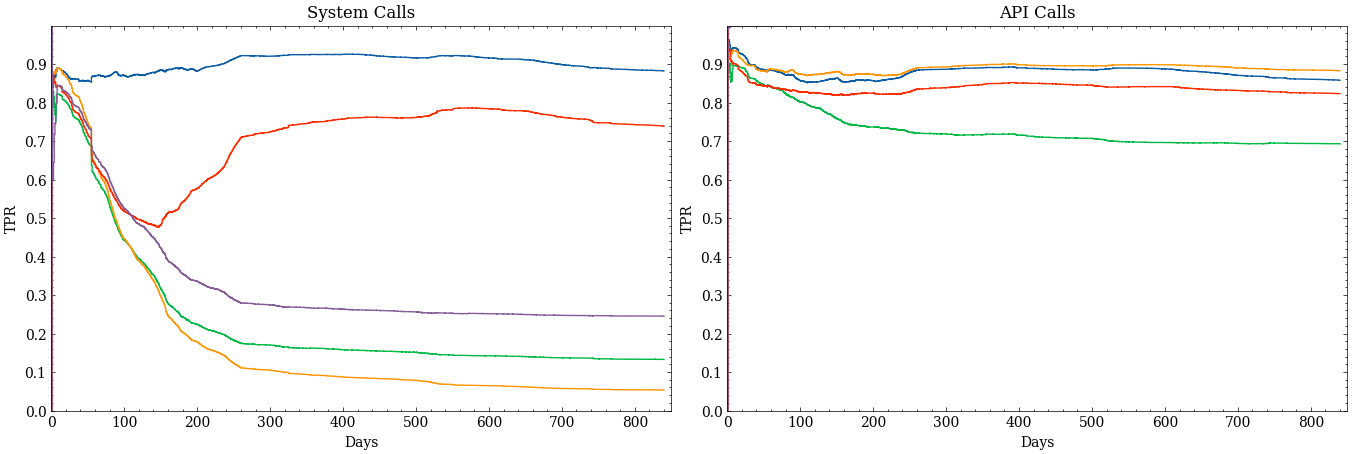}
         \caption{TPR}
         \label{fig:adt}
     \end{subfigure}

        \caption{Active learning framework results on OL models trained on dynamic features}
        \label{fig:activedynamic}
\end{figure}

\begin{figure}
     \centering
         \begin{subfigure}{\textwidth}
     \centering
         \includegraphics[width=\columnwidth]{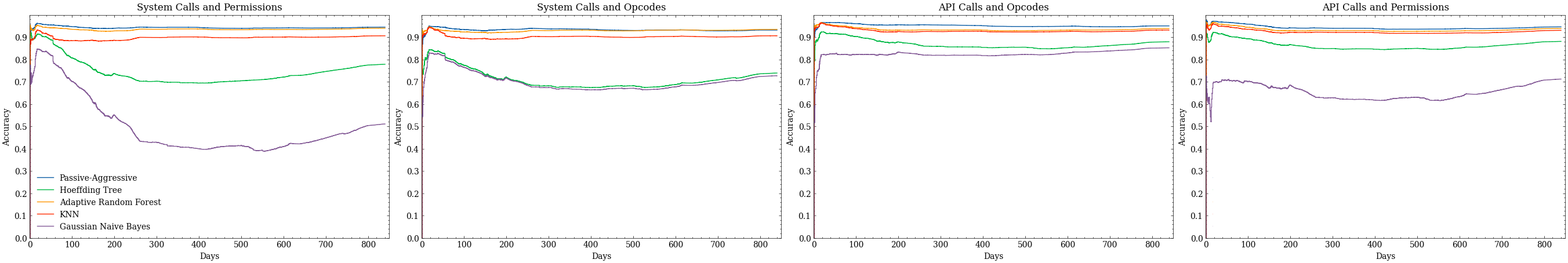}
         \caption{Accuracy}
         \label{fig:aha}
     \end{subfigure}
    
     \begin{subfigure}{\textwidth}
         \centering
         \includegraphics[width=\columnwidth]{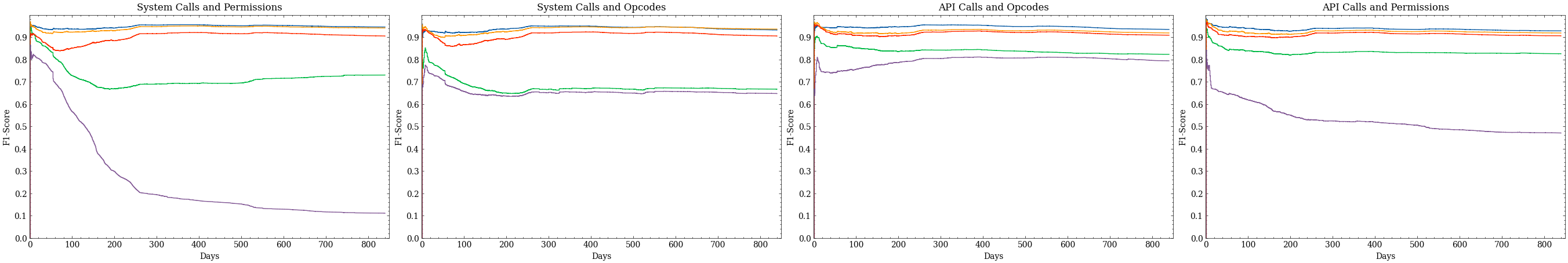}
         \caption{F1-Score}
         \label{fig:ahf}
     \end{subfigure}
 
     \begin{subfigure}{\textwidth}
         \centering
         \includegraphics[width=\columnwidth]{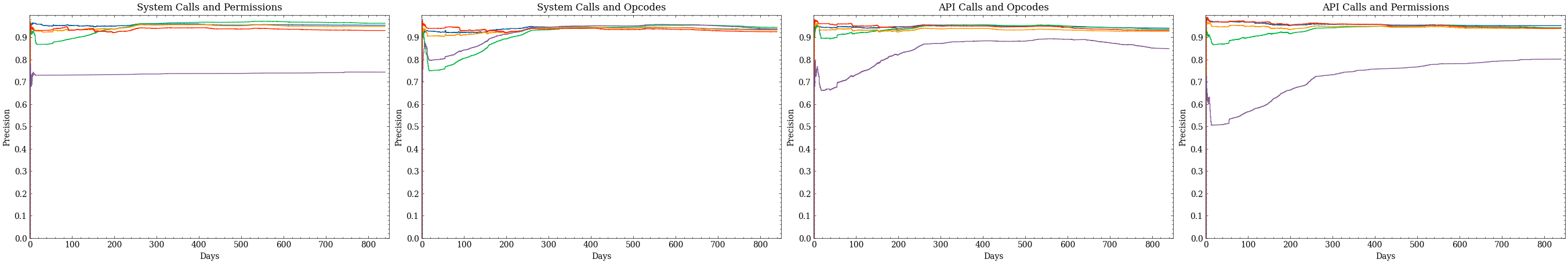}
         \caption{Precision}
         \label{fig:ahp}
     \end{subfigure}

     \begin{subfigure}{\textwidth}
         \centering
         \includegraphics[width=\columnwidth]{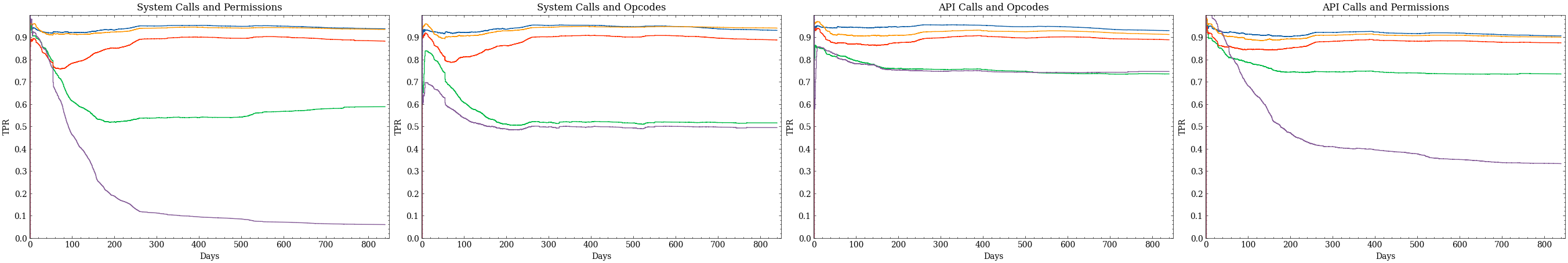}
         \caption{TPR}
         \label{fig:aht}
     \end{subfigure}

        \caption{Active learning framework results on OL models trained on hybrid features}
        \label{fig:activehybrid}
\end{figure}

In this section, we report the results of our active learning framework evaluated within the delayed progressive validation scenario. Table \ref{tab:a_table} summarises the results, showing the average metrics for the best performing class of OL model when different feature sets are used. The table also shows the proportion of the training data for which a model requests labels. Figures \ref{fig:activestatic}--\ref{fig:activehybrid} plot detailed results for the 5 types of OL model when static, dynamic and hybrid feature sets are used.

\begin{table}[t]
    \caption{Summary of performance of the active learning framework under the delayed progressive validation scenario. For each feature set, the mean and standard deviations of metrics of the best-performing OL model are shown. Horizontal rules separate the static, dynamic and hybrid feature sets. The final column indicates the proportion of labels requested.}
    \label{tab:a_table}
    \centering
    \resizebox{\columnwidth}{!}{%
        \begin{tabular}{@{}llllllllr@{}}
            \toprule
            \textbf{Feature}             & \textbf{OL Model} & \textbf{Accuracy}  & \textbf{F1-Score} & \textbf{Precision}& \textbf{TPR} & \textbf{Labels} & \textbf{Drifts}  \\
            \midrule
            Permissions                  & PA                & 0.924$ \pm 0.016$      & 0.927$ \pm 0.015$  & 0.947$ \pm 0.018$         & 0.908$ \pm 0.017$                       & 31\%  & 15                                      \\
            API Calls & PA & \textbf{0.960}$ \pm 0.011$   & \textbf{0.963}$ \pm 0.013$  & \textbf{0.961}$ \pm 0.015$  & \textbf{0.964}$ \pm 0.014$   & 34\% & 12 
            \\
            Opcodes                      & PA                & 0.924$ \pm 0.022$       & 0.929$ \pm 0.024$  &  0.923$ \pm 0.022$        & 0.936$ \pm 0.026$                        & 32\%        & 12                                    \\
            \midrule
            System Calls                 &  PA                & 0.867$ \pm 0.023$     & 0.882$ \pm 0.027$  & 0.862$ \pm 0.026$          & 0.897$ \pm 0.031$                      & 30\%           & 17                                    \\
            
            Dynamic API Calls                & PA                & 0.901$ \pm 0.022$      & 0.897$ \pm 0.026$  & 0.913$ \pm 0.021$         & 0.880$ \pm 0.028$                      & 27\%          & 16                                \\
            \midrule
            System Calls and Permissions & PA                & 0.951$ \pm 0.018$     & 0.947$ \pm 0.021$  & 0.953$ \pm 0.018$          & 0.940$ \pm 0.024$                       & 29\%                & 13                       \\
            System Calls and Opcodes     &  RF                & 0.940$ \pm 0.012$   & 0.937$ \pm 0.018$  & 0.935$ \pm 0.016$            & 0.938$ \pm 0.022$                       & 28\%                & 12       \\             
                      Dynamic API Calls and Opcodes     &  RF                & 0.953$ \pm 0.013$   & 0.947$ \pm 0.016$  & 0.950$ \pm 0.017$            & 0.946$ \pm 0.014$                       & \textbf{24\%}       & 16              \\  
            Dynamic API Calls and Permissions & PA                & 0.954$ \pm 0.014$     & 0.936$ \pm 0.013$  & 0.956$ \pm 0.013$          & 0.917$ \pm 0.016$                        & 25\%                 & 14                      \\
        
            \bottomrule
        \end{tabular}%
    }
\end{table}

Notably, the results of the active learning framework are much more similar to standard-trained OL models evaluated under progressive validation than those evaluated under delayed progressive validation. The best models achieve accuracies of about 96\%, which is only 1\% less than that achieved by standard-trained OL models in a progressive validation scenario, and 5\% better than that achieved by standard-trained OL models in a delayed progressive validation scenario. From the plots, it is clear that there is no longer a pronounced dip in accuracy once incremental learning starts, suggesting that active learning compensates for the concept drift that occurs between application release and the updating of the model.

With active learning, static API calls appear to be the most effective feature set to use, at least in terms of achieving high accuracy and low variance. However, another consideration here is the number of labelling requests. In this respect, static API calls are the most costly feature, with labels requested for 34\% of the training data. The lowest levels of label requests are seen with models trained on hybrid feature sets, with the two dynamic API call hybrid feature sets requiring labels only around 25\% of the time. This is a significant difference, and may compensate for the marginal loss in accuracy.

In other respects, too, the results for active learning resemble those for standard-trained OL evaluated within the progressive evaluation context. Notably, PA is again the dominant model, and models trained on permissions return to a baseline of around 92\% accuracy.




\section{Discussion}

\begin{table}[t]
	\caption{Overview of results.}
	\label{tab:oleval}
	\centering
	\resizebox{\columnwidth}{!}{%
		\begin{tabular}{@{}llllllllll@{}}
			\toprule
			\multicolumn{1}{c}{\textbf{ }}   & \multicolumn{3}{c}{\textbf{Progressive   Validation}}                                                                                 & \multicolumn{3}{c}{\textbf{Active Learning}}                                                           & \multicolumn{3}{c}{\textbf{Delayed Progressive Validation}}                                            \\
			\textbf{Feature}             & \textbf{Accuracy} & \textbf{TPR}  & \textbf{Labels} & \textbf{Accuracy} & \textbf{TPR}  & \textbf{Labels} & \textbf{Accuracy} & \textbf{TPR}  & \textbf{Labels} \\
			\midrule
			Permissions                  & 0.928$ \pm 0.016$              & 0.916$ \pm 0.017$                      & 100\%                         & 0.924$ \pm 0.016$              & 0.908$ \pm 0.017$                     & 31\%                        & 0.888$ \pm 0.040$              & 0.848$ \pm 0.038$                   & 100\%                         \\
			API Calls & 0.971$ \pm 0.007$   & 0.978$ \pm 0.007$   & 100\% & 0.960$ \pm 0.011$   & 0.964$ \pm 0.014$   & 34\% & 0.783$ \pm 0.015$   & 0.920$ \pm 0.024$   & 100\% 
			\\
			Opcodes                      & 0.952$ \pm 0.029$              & 0.958$ \pm 0.031$                      & 100\%                         & 0.924$ \pm 0.022$              & 0.936$ \pm 0.026$                       & 32\%                          & 0.839$ \pm 0.047$              & 0.798$ \pm 0.065$                      & 100\%                         \\
			\midrule
			System Calls                 & 0.911$ \pm 0.031$              & 0.932$ \pm 0.042$                        & 100\%                         & 0.867$ \pm 0.023$              & 0.897$ \pm 0.031$                       & 30\%                          & 0.782$ \pm 0.034$              & 0.803$ \pm 0.063$                       & 100\%                         \\
			
			Dynamic API Calls               & 0.912$ \pm 0.013$              & 0.889$ \pm 0.025$                       & 100\%                         & 0.901$ \pm 0.022$              & 0.880$ \pm 0.028$                       & 27\%                          & 0.832$ \pm 0.037$              & 0.802$ \pm 0.059$                     & 100\%                         \\
			\midrule
			System Calls and Permissions & 0.952$ \pm 0.019$              & 0.952$ \pm 0.025$                       & 100\%                         & 0.951$ \pm 0.018$              & 0.947$ \pm 0.024$                       & 29\%                          &  0.872$ \pm 0.036$              & 0.941$ \pm 0.026$                      & 100\%                         \\
			System Calls and Opcodes     & 0.958$ \pm 0.027$              & 0.952$ \pm 0.028$                       & 100\%                         & 0.940$ \pm 0.012$              & 0.938$ \pm 0.022$                        & 28\%                        & 0.853$ \pm 0.057$              & 0.825$ \pm 0.033$                      & 100\%      \\
   		Dynamic API Calls and Opcodes     & 0.959$ \pm 0.025$              & 0.954$ \pm 0.025$                       & 100\%                         & 0.953$ \pm 0.013$              & 0.946$ \pm 0.014$                        & 24\%                         & 0.832$ \pm 0.077$              & 0.714$ \pm 0.034$                      & 100\%  \\   
			Dynamic API Calls and Permissions & 0.947$ \pm 0.013$              & 0.924$ \pm 0.015$                       & 100\%                         & 0.954$ \pm 0.014$              & 0.917$ \pm 0.016$                       & 25\%                          & 0.908$ \pm 0.032$              & 0.870$ \pm 0.052$                      & 100\%                         \\
	
			\bottomrule
		\end{tabular}%
	}
\end{table}

Table \ref{tab:oleval} provides a direct comparison between the models trained in the three experimental sections. This further emphasises the fact the results for active learning are close to the ideal baseline achieved when OL models are trained under the assumption that labels are instantly available, even though they are trained with a labelling delay. This is very encouraging, since it suggests that the use of active learning overcomes the significant deficit in performance that normally occurs when a labelling delay is introduced.

It does raise the question of where this resilience comes from, since there are two parts to the active learning framework: the selective use of training data, and the periodic retraining after concept drift has been detected. Periodic retraining means that the models can offload the baggage of historical information that is no longer relevant, and this by itself would seem to give an advantage over standard OL. However, if this was the only factor, then we would expect to see a periodic pattern of decline and improvement in the performance plots, which is not evident. This suggests that selective training (the core of active learning) is also important, and that selective use of training data causes the models to better react to change, since they are generally being trained on data that does not resemble that which they have seen before (assuming such data would have higher confidence). This is also evident in the considerably lower number of drifts detected compared to delayed progressive validation on each model, while maintaining similar levels to progressive validation.

We have discussed the influence of models, features and training regime, but it is also important to reflect on the fact that there are trade-offs in the design of any real world system. We have noted several: the time and effort required to extract features, the time and computational resources required to build models, and the amount of labelled data required. The best choice of machine learning approach is, to some extent, influenced by the relative weightings placed on these three factors.

The extraction of dynamic features requires that an application is executed for a certain period of time within a simulator. This process took about 2-3 minutes for each of the applications in our data set, where on most days there were up to around 400 applications released. On the machine we used, it was generally feasible to carry out dynamic analysis of only a couple of applications in parallel. This means that, in practice, assuming the same computational resources, a sizeable part of each day would be required just to extract dynamic features. This limits the practical utility of using dynamic and hybrid feature sets for training online learning models. So, whilst there does appear to be some benefit in terms of accuracy to using these feature sets in certain scenarios (notably, delayed progressive validation when active learning is not used), it may not work as a practical solution.

The time and computational resources required to build models is less of a concern. In our experiments, most models could be incrementally updated in seconds for each new application. Retraining from seed data following periods of concept drift was also easily manageable using our modest computational resources. The only real issue, as noted above, was dealing with large feature sets, meaning in particular that -- although the best feature in terms of accuracy --- building models from static API feature sets can be problematic. However, this could potentially be addressed using a better-equipped production environment.

From our perspective at least, obtaining labelled data is the most expensive of these processes. Currently, with a free-to-use public license, the VirusTotal API is limited to 500 requests per day at a rate of 4 per minute. The reduction in labelling requests that results from using active learning is consequently an important part of making this framework practical, and comes as a significant benefit in addition to the improvement in accuracy. However, it should be noted that higher limits are available for commercial licences, so this may be less of an issue in a non-academic context.

\section{Conclusions}

The interplay between attackers and defenders means that malware is very likely to display concept drift over time, with the underlying basis of malware attacks changing as anti-malware products respond to these attacks. Given this, it is perhaps surprising that most prior work on Android malware detection has treated it as a batch learning problem rather than an online learning problem. Those who have treated it as an online learning problem have typically evaluated their methodologies under the assumption that the label for an application is known immediately upon its release. In practice, this is unlikely, and it will only become clear later on whether an application is malware or benign.

Using a large contemporary dataset, we have shown that this labelling delay significantly impedes the predictive performance of standard online learning models trained to recognise malware, due to the concept drift that occurs in the period between application release and application labelling. Models trained and evaluated under the assumption that labels are immediately available achieve accuracies of up to 97\%. When realistic labelling delays are introduced, the best accuracies drop to around 91\%. Furthermore, observations about the best choice of model do not remain valid when a labelling delay is introduced.

In this paper, we have introduced a novel active learning framework, and have shown that it largely compensates for the performance reduction observed when realistic labelling delays are introduced, resulting in models with accuracies of up to 96\%. Importantly, it also reduces the amount of labelled data required to train models, with only 24-34\% of the training data used, depending on the chosen model. This is important because labelling is costly. Our framework consists of two main parts: the selective use of training data during incremental learning, and the periodic retraining of models from seed data when concept drift is detected in the existing model. Both of these components appear to be important.

We also carried out an in-depth investigation of the influence of model and feature choice within an online learning context. Of the five models we investigated, passive-aggressive classifiers were generally the best performing, though adaptive random forests and adaptive KNNs also performed well. Naive Bayes performed very poorly. Of the nine feature sets we used, measures of API call usage obtained from static analysis generally led to the best performing models. However, the high-dimensionality of this feature set was problematic. We observed around a 4\% penalty to using the more easily extracted, and far less numerous, permissions and opcode features.

Features extracted using dynamic analysis were generally less useful, at least by themselves. However, when combined with static features, they led to the best models when non-active learning models were evaluated within a delayed labelling context. When active learning was used, these hybrid feature combinations required the least training data, and were the best performing after static API calls. However, there is a considerable cost to doing dynamic analysis, and this may limit the practical deployment of these hybrid feature sets.

Future Android operating system updates may introduce new permissions and API calls. As a result, future works may include dynamic feature sets to train the online learning models. This may be using the full new feature sets or using feature selection algorithms to select the most relevant features in intervals. Using an ensemble of online learning models in the future can also result in robust classifiers.

\appendix
\section{Results of Standard Models} \label{appendix}

\begin{figure}[H]
     \centering
         \begin{subfigure}{\textwidth}
     \centering
         \includegraphics[width=\columnwidth]{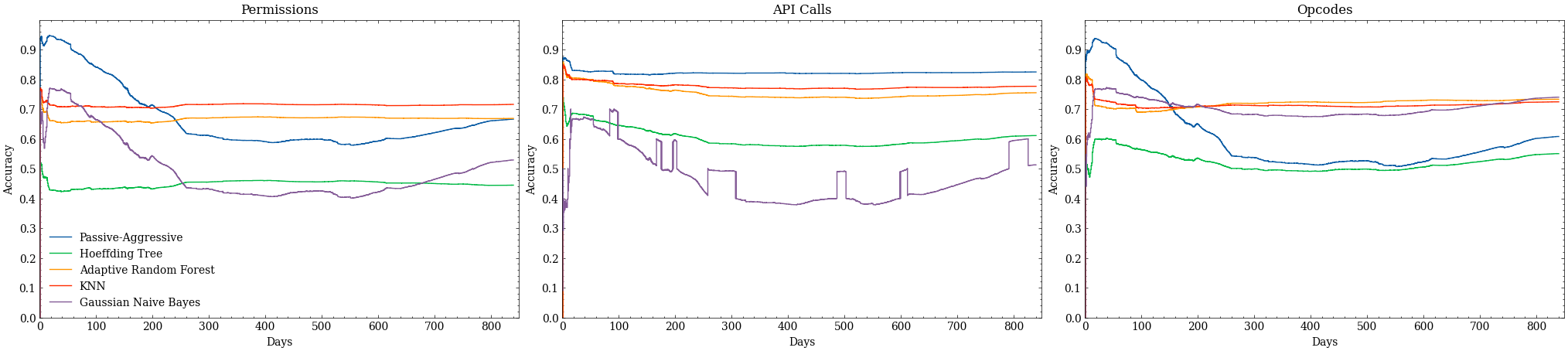}
         \caption{Accuracy}
         \label{fig:ssa}
     \end{subfigure}
    
     \begin{subfigure}{\textwidth}
         \centering
         \includegraphics[width=\columnwidth]{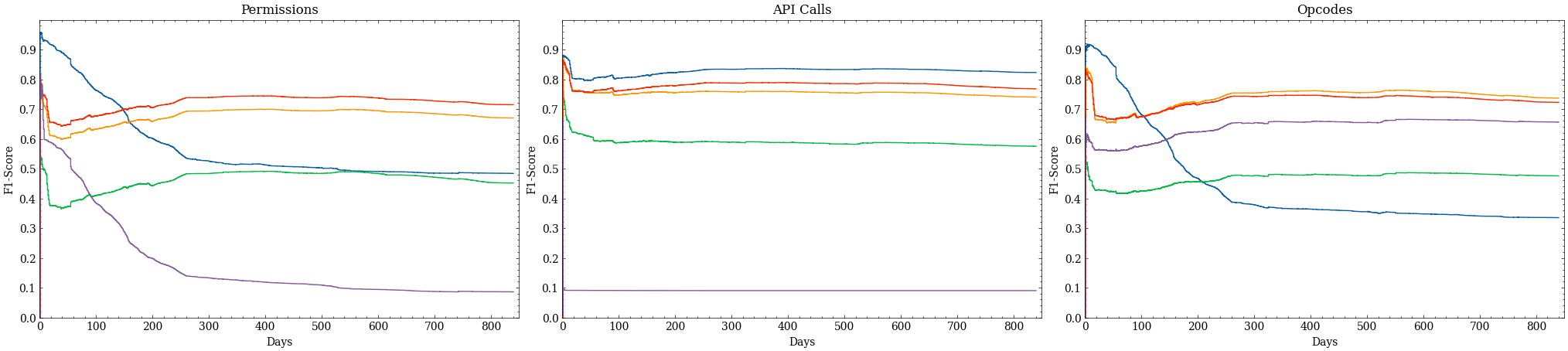}
         \caption{F1-Score}
         \label{fig:ssf}
     \end{subfigure}
 
     \begin{subfigure}{\textwidth}
         \centering
         \includegraphics[width=\columnwidth]{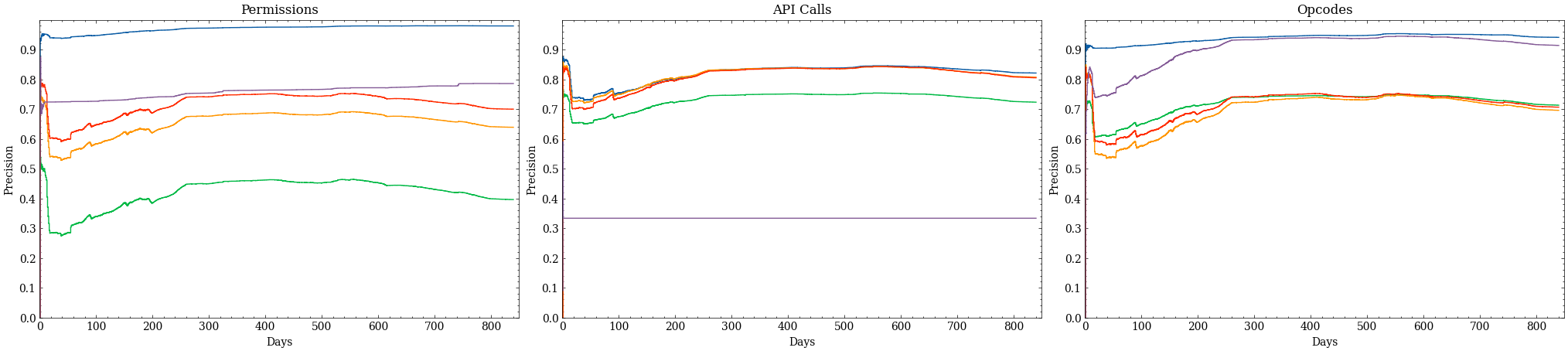}
         \caption{Precision}
         \label{fig:ssp}
     \end{subfigure}

     \begin{subfigure}{\textwidth}
         \centering
         \includegraphics[width=\columnwidth]{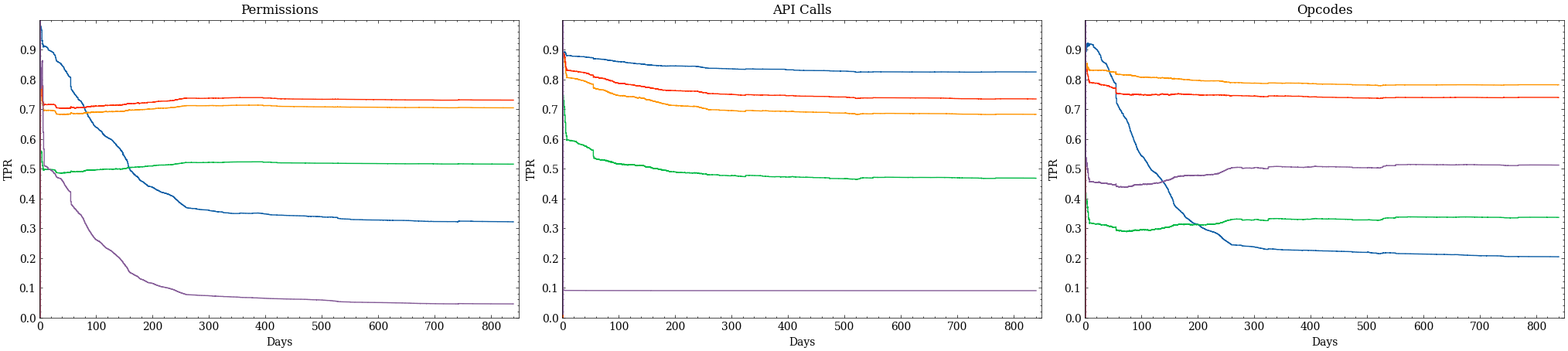}
         \caption{TPR}
         \label{fig:sst}
     \end{subfigure}

        \caption{Results of OL models trained on static features with only initial seed data and no further updates}
        \label{fig:standardstatic}
\end{figure}

\begin{figure}[H]
     \centering
     \begin{subfigure}{\textwidth}
     \centering
         \includegraphics[width=\columnwidth]{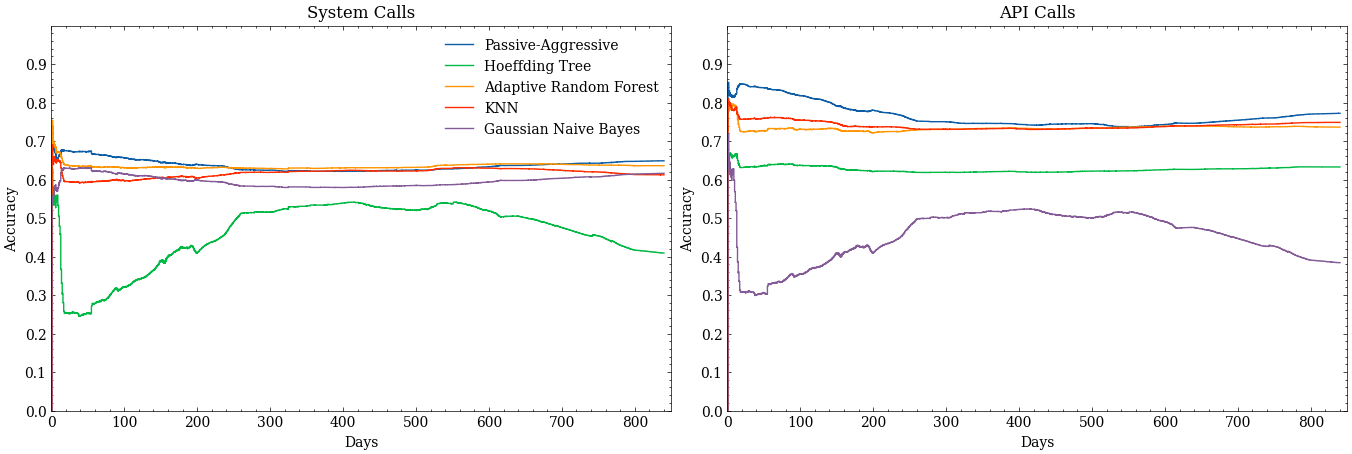}
         \caption{Accuracy}
         \label{fig:sda}
     \end{subfigure}
    
     \begin{subfigure}{\textwidth}
         \centering
         \includegraphics[width=\columnwidth]{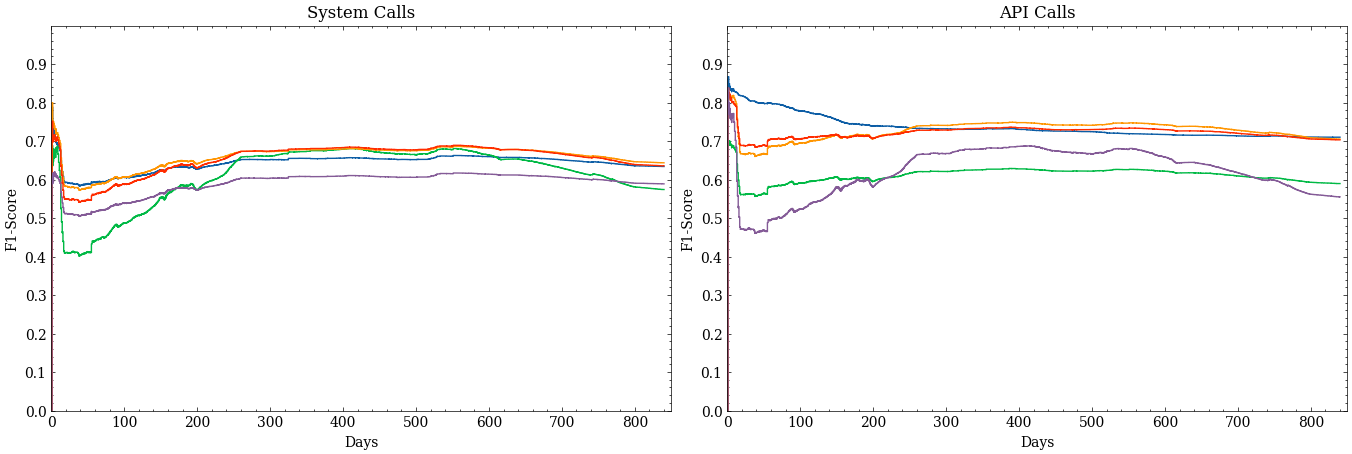}
         \caption{F1-Score}
         \label{fig:sdf}
     \end{subfigure}
 
     \begin{subfigure}{\textwidth}
         \centering
         \includegraphics[width=\columnwidth]{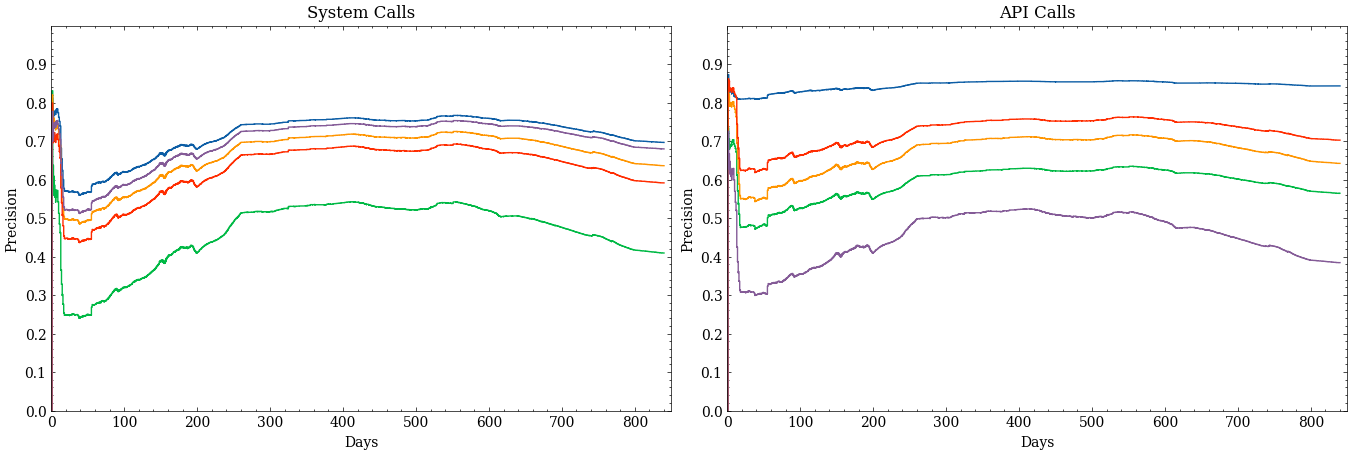}
         \caption{Precision}
         \label{fig:sdp}
     \end{subfigure}

     \begin{subfigure}{\textwidth}
         \centering
         \includegraphics[width=\columnwidth]{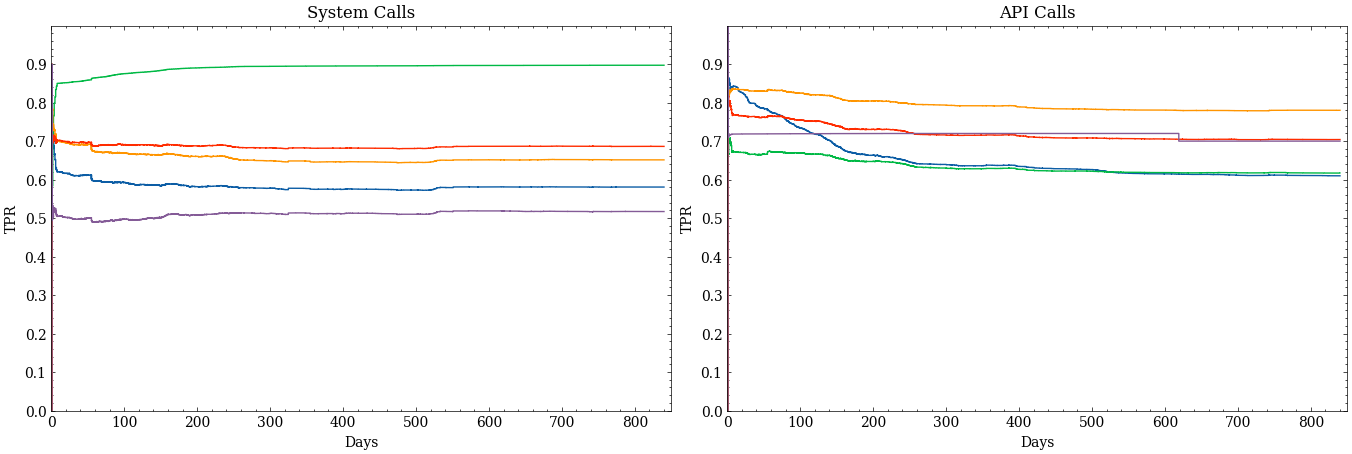}
         \caption{TPR}
         \label{fig:sdt}
     \end{subfigure}

        \caption{Results of OL models trained on static features with only initial seed data and no further updates}
        \label{fig:standarddynamic}
\end{figure}

\begin{figure}[H]
     \centering
         \begin{subfigure}{\textwidth}
     \centering
         \includegraphics[width=\columnwidth]{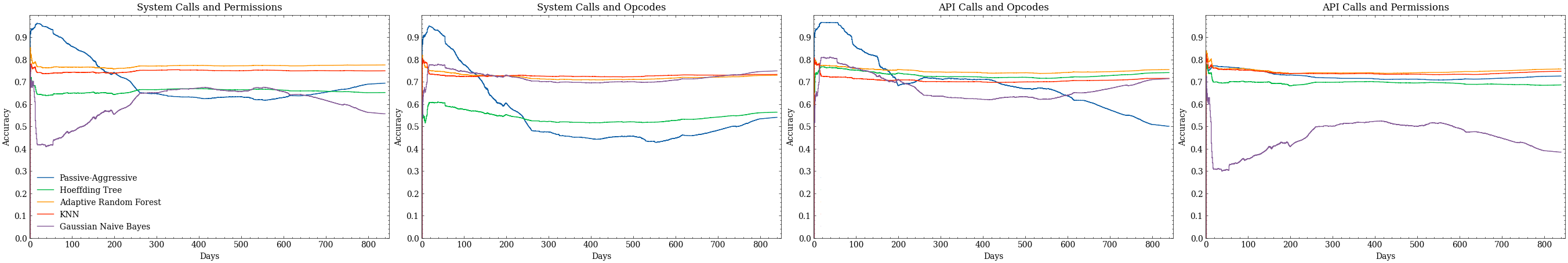}
         \caption{Accuracy}
         \label{fig:sha}
     \end{subfigure}
    
     \begin{subfigure}{\textwidth}
         \centering
         \includegraphics[width=\columnwidth]{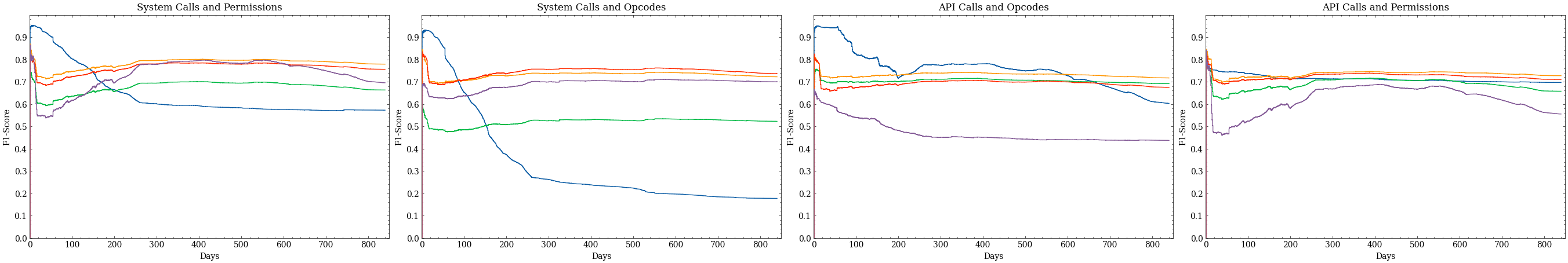}
         \caption{F1-Score}
         \label{fig:shf}
     \end{subfigure}
 
     \begin{subfigure}{\textwidth}
         \centering
         \includegraphics[width=\columnwidth]{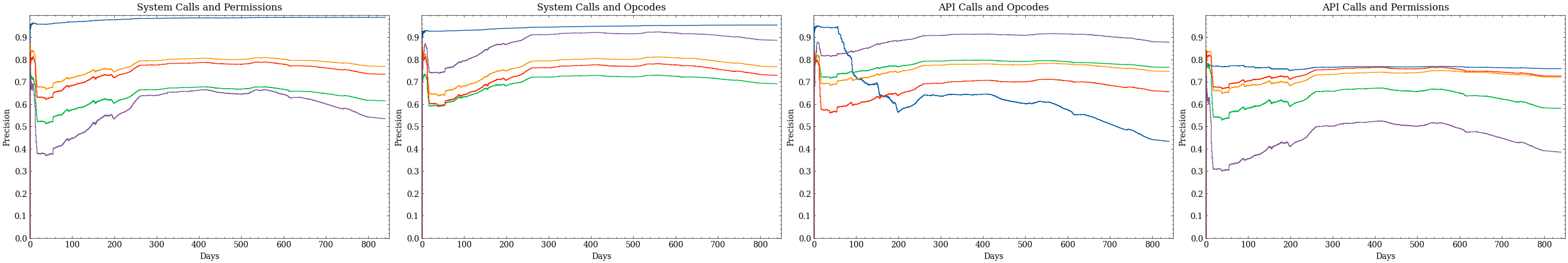}
         \caption{Precision}
         \label{fig:shp}
     \end{subfigure}

     \begin{subfigure}{\textwidth}
         \centering
         \includegraphics[width=\columnwidth]{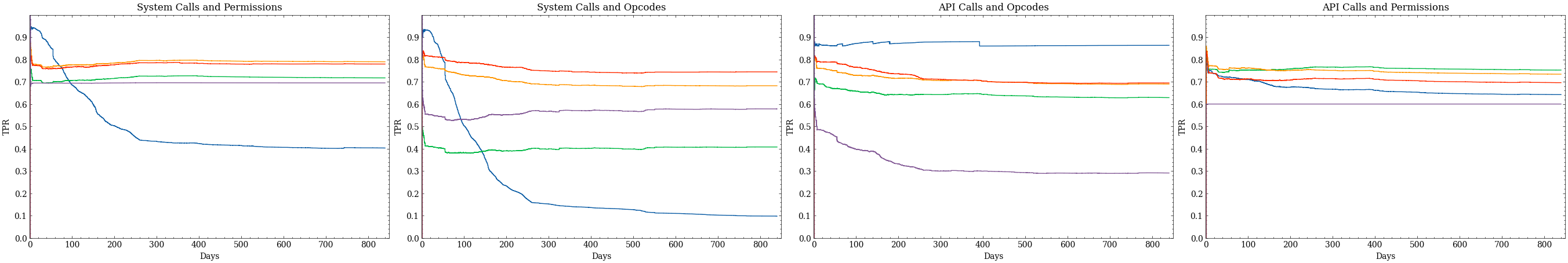}
         \caption{TPR}
         \label{fig:sht}
     \end{subfigure}

        \caption{Results of OL models trained on hybrid features with only initial seed data and no further updates}
        \label{fig:standardhybrid}
\end{figure}

We carried out another simulation in which the OL models were only trained with seed data and thereafter evaluated and not updated with new data. Figures \ref{fig:standardstatic}--\ref{fig:standardhybrid} plot detailed results for the 5 types of OL model when static, dynamic, and hybrid feature sets are used. All the models show a decline in accuracy and F1-score after a small period initially. 

Models trained using static features showed a constant decline in accuracy with new applications. The models also showed high precision rates and low TPR suggests that the models may be overly conservative in predicting positive instances, resulting in a low TPR while maintaining a low false positive rate.  

The performance of dynamic features was comparable to that of static features, with a little increase in metrics midway through and a progressive decline trend toward the end.

While hybrid features showed stable F1-scores, their accuracy significantly dropped. The models also showed high precision rates but low TPR, which suggested the model is extremely cautious in classifying applications as malware.

A considerable decline in performance can be seen and proves the use of OL helps in maintaining the performance of the models over time. 

\def\UrlBreaks{\do\/\do-}
\bibliographystyle{elsarticle-harv}
\bibliography{library}

\begin{thebibliography}{39}
\expandafter\ifx\csname natexlab\endcsname\relax\def\natexlab#1{#1}\fi
\providecommand{\url}[1]{\texttt{#1}}
\providecommand{\href}[2]{#2}
\providecommand{\path}[1]{#1}
\providecommand{\DOIprefix}{doi:}
\providecommand{\ArXivprefix}{arXiv:}
\providecommand{\URLprefix}{URL: }
\providecommand{\Pubmedprefix}{pmid:}
\providecommand{\doi}[1]{\href{http://dx.doi.org/#1}{\path{#1}}}
\providecommand{\Pubmed}[1]{\href{pmid:#1}{\path{#1}}}
\providecommand{\bibinfo}[2]{#2}
\ifx\xfnm\relax \def\xfnm[#1]{\unskip,\space#1}\fi
\bibitem[{Arp et~al.(2014)Arp, Spreitzenbarth, H{\"{u}}bner, Gascon and
  Rieck}]{Arp2014}
\bibinfo{author}{Arp, D.}, \bibinfo{author}{Spreitzenbarth, M.},
  \bibinfo{author}{H{\"{u}}bner, M.}, \bibinfo{author}{Gascon, H.},
  \bibinfo{author}{Rieck, K.}, \bibinfo{year}{2014}.
\newblock \bibinfo{title}{{Drebin: Effective and Explainable Detection of
  Android Malware in Your Pocket}}, in: \bibinfo{booktitle}{Network and
  Distributed System Security Symposium (NDSS)}.
\newblock \DOIprefix\doi{10.14722/ndss.2014.23247}.
\bibitem[{Bai et~al.(2020)Bai, Xie, Di and Ye}]{Bai2020}
\bibinfo{author}{Bai, H.}, \bibinfo{author}{Xie, N.}, \bibinfo{author}{Di, X.},
  \bibinfo{author}{Ye, Q.}, \bibinfo{year}{2020}.
\newblock \bibinfo{title}{{FAMD: A fast multifeature android malware detection
  framework, design, and implementation}}.
\newblock \bibinfo{journal}{IEEE Access} \bibinfo{volume}{8},
  \bibinfo{pages}{194729--194740}.
\newblock \DOIprefix\doi{10.1109/ACCESS.2020.3033026}.
\bibitem[{Bifet and Gavalda(2007)}]{bifet2007a}
\bibinfo{author}{Bifet, A.}, \bibinfo{author}{Gavalda, R.},
  \bibinfo{year}{2007}.
\newblock \bibinfo{title}{Learning from time-changing data with adaptive
  windowing}, in: \bibinfo{booktitle}{Proceedings of the 2007 SIAM
  international conference on data mining}, \bibinfo{organization}{SIAM}. pp.
  \bibinfo{pages}{443--448}.
\bibitem[{Bifet et~al.(2010)Bifet, Holmes, Kirkby and Pfahringer}]{moa}
\bibinfo{author}{Bifet, A.}, \bibinfo{author}{Holmes, G.},
  \bibinfo{author}{Kirkby, R.}, \bibinfo{author}{Pfahringer, B.},
  \bibinfo{year}{2010}.
\newblock \bibinfo{title}{{MOA:} massive online analysis}.
\newblock \bibinfo{journal}{J. Mach. Learn. Res.} \bibinfo{volume}{11},
  \bibinfo{pages}{1601--1604}.
\newblock \URLprefix \url{http://portal.acm.org/citation.cfm?id=1859903}.
\bibitem[{Ceschin et~al.(2023)Ceschin, Botacin, Gomes, Pinag{\'e}, Oliveira and
  Gr{\'e}gio}]{ceschin2023}
\bibinfo{author}{Ceschin, F.}, \bibinfo{author}{Botacin, M.},
  \bibinfo{author}{Gomes, H.M.}, \bibinfo{author}{Pinag{\'e}, F.},
  \bibinfo{author}{Oliveira, L.S.}, \bibinfo{author}{Gr{\'e}gio, A.},
  \bibinfo{year}{2023}.
\newblock \bibinfo{title}{Fast \& furious: On the modelling of malware
  detection as an evolving data stream}.
\newblock \bibinfo{journal}{Expert Systems with Applications}
  \bibinfo{volume}{212}, \bibinfo{pages}{118590}.
\bibitem[{Crammer et~al.(2006)Crammer, Dekel, Keshet, Shalev-Shwartz and
  Singer}]{crammer2006}
\bibinfo{author}{Crammer, K.}, \bibinfo{author}{Dekel, O.},
  \bibinfo{author}{Keshet, J.}, \bibinfo{author}{Shalev-Shwartz, S.},
  \bibinfo{author}{Singer, Y.}, \bibinfo{year}{2006}.
\newblock \bibinfo{title}{Online passive aggressive algorithms} .
\bibitem[{Data(2022)}]{malwareStat}
\bibinfo{author}{Data, G.}, \bibinfo{year}{2022}.
\newblock \bibinfo{title}{G data mobile security report: Conflict in ukraine
  causes decline in malicious android apps}.
\bibitem[{Gomes et~al.(2017)Gomes, Bifet, Read, Barddal, Enembreck, Pfharinger,
  Holmes and Abdessalem}]{Gomes2017}
\bibinfo{author}{Gomes, H.M.}, \bibinfo{author}{Bifet, A.},
  \bibinfo{author}{Read, J.}, \bibinfo{author}{Barddal, J.P.},
  \bibinfo{author}{Enembreck, F.}, \bibinfo{author}{Pfharinger, B.},
  \bibinfo{author}{Holmes, G.}, \bibinfo{author}{Abdessalem, T.},
  \bibinfo{year}{2017}.
\newblock \bibinfo{title}{Adaptive random forests for evolving data stream
  classification}.
\newblock \bibinfo{journal}{Machine Learning} \bibinfo{volume}{106},
  \bibinfo{pages}{1469--1495}.
\bibitem[{Guerra-Manzanares and Bahsi(2022)}]{GUERRAMANZANARES2022}
\bibinfo{author}{Guerra-Manzanares, A.}, \bibinfo{author}{Bahsi, H.},
  \bibinfo{year}{2022}.
\newblock \bibinfo{title}{On the relativity of time: Implications and
  challenges of data drift on long-term effective android malware detection}.
\newblock \bibinfo{journal}{Computers \& Security} \bibinfo{volume}{122},
  \bibinfo{pages}{102835}.
\bibitem[{Hou et~al.(2017)Hou, Ye, Song and Abdulhayoglu}]{Hou2017}
\bibinfo{author}{Hou, S.}, \bibinfo{author}{Ye, Y.}, \bibinfo{author}{Song,
  Y.}, \bibinfo{author}{Abdulhayoglu, M.}, \bibinfo{year}{2017}.
\newblock \bibinfo{title}{{Hin droid: An intelligent Android Malware detection
  system based on structured heterogeneous information network}}.
\newblock \bibinfo{journal}{Proceedings of the ACM SIGKDD International
  Conference on Knowledge Discovery and Data Mining} \bibinfo{volume}{Part
  F1296}, \bibinfo{pages}{1507--1516}.
\newblock \DOIprefix\doi{10.1145/3097983.3098026}.
\bibitem[{Kan et~al.(2021)Kan, Pendlebury, Pierazzi and Cavallaro}]{kan2021}
\bibinfo{author}{Kan, Z.}, \bibinfo{author}{Pendlebury, F.},
  \bibinfo{author}{Pierazzi, F.}, \bibinfo{author}{Cavallaro, L.},
  \bibinfo{year}{2021}.
\newblock \bibinfo{title}{Investigating labelless drift adaptation for malware
  detection}, in: \bibinfo{booktitle}{Proceedings of the 14th ACM Workshop on
  Artificial Intelligence and Security}, pp. \bibinfo{pages}{123--134}.
\bibitem[{Kandukuru and Sharma(2017)}]{Kandukuru2017}
\bibinfo{author}{Kandukuru, S.}, \bibinfo{author}{Sharma, R.M.},
  \bibinfo{year}{2017}.
\newblock \bibinfo{title}{Android malicious application detection using
  permission vector and network traffic analysis}, in: \bibinfo{booktitle}{2017
  2nd International Conference for Convergence in Technology (I2CT)}, pp.
  \bibinfo{pages}{1126--1132}.
\newblock \DOIprefix\doi{10.1109/I2CT.2017.8226303}.
\bibitem[{Kang et~al.(2016)Kang, Yerima, McLaughlin and Sezer}]{Kang2016}
\bibinfo{author}{Kang, B.J.}, \bibinfo{author}{Yerima, S.Y.},
  \bibinfo{author}{McLaughlin, K.}, \bibinfo{author}{Sezer, S.},
  \bibinfo{year}{2016}.
\newblock \bibinfo{title}{{N-opcode analysis for android malware classification
  and categorization}}.
\newblock \bibinfo{journal}{2016 International Conference on Cyber Security and
  Protection of Digital Services, Cyber Security 2016} ,
  \bibinfo{pages}{13--14}\DOIprefix\doi{10.1109/CyberSecPODS.2016.7502343}.
\bibitem[{Li et~al.(2017)Li, Wang, Li, Wang, Wang and Xue}]{Li2017}
\bibinfo{author}{Li, D.}, \bibinfo{author}{Wang, Z.}, \bibinfo{author}{Li, L.},
  \bibinfo{author}{Wang, Z.}, \bibinfo{author}{Wang, Y.}, \bibinfo{author}{Xue,
  Y.}, \bibinfo{year}{2017}.
\newblock \bibinfo{title}{{FgDetector: Fine-Grained Android Malware
  Detection}}.
\newblock \bibinfo{journal}{Proceedings - 2017 IEEE 2nd International
  Conference on Data Science in Cyberspace, DSC 2017} ,
  \bibinfo{pages}{311--318}\DOIprefix\doi{10.1109/DSC.2017.13}.
\bibitem[{Lindorfer et~al.(2015)Lindorfer, Neugschwandtner and
  Platzer}]{Lindorfer2015}
\bibinfo{author}{Lindorfer, M.}, \bibinfo{author}{Neugschwandtner, M.},
  \bibinfo{author}{Platzer, C.}, \bibinfo{year}{2015}.
\newblock \bibinfo{title}{{MARVIN: Efficient and Comprehensive Mobile App
  Classification through Static and Dynamic Analysis}}.
\newblock \bibinfo{journal}{Proceedings - International Computer Software and
  Applications Conference} \bibinfo{volume}{2}, \bibinfo{pages}{422--433}.
\newblock \DOIprefix\doi{10.1109/COMPSAC.2015.103}.
\bibitem[{Ma et~al.(2019)Ma, Ge, Liu, Zhao and Ma}]{Ma2019}
\bibinfo{author}{Ma, Z.}, \bibinfo{author}{Ge, H.}, \bibinfo{author}{Liu, Y.},
  \bibinfo{author}{Zhao, M.}, \bibinfo{author}{Ma, J.}, \bibinfo{year}{2019}.
\newblock \bibinfo{title}{{A Combination Method for Android Malware Detection
  Based on Control Flow Graphs and Machine Learning Algorithms}}.
\newblock \bibinfo{journal}{IEEE Access} \bibinfo{volume}{7},
  \bibinfo{pages}{21235--21245}.
\newblock \DOIprefix\doi{10.1109/ACCESS.2019.2896003}.
\bibitem[{Mirzaei et~al.(2019)Mirzaei, de~Fuentes, Tapiador and
  Gonzalez-Manzano}]{Mirzaei2019}
\bibinfo{author}{Mirzaei, O.}, \bibinfo{author}{de~Fuentes, J.M.},
  \bibinfo{author}{Tapiador, J.}, \bibinfo{author}{Gonzalez-Manzano, L.},
  \bibinfo{year}{2019}.
\newblock \bibinfo{title}{{ANDRODET: An adaptive Android obfuscation
  detector}}.
\newblock \bibinfo{journal}{Future Generation Computer Systems}
  \bibinfo{volume}{90}, \bibinfo{pages}{240--261}.
\newblock \DOIprefix\doi{10.1016/j.future.2018.07.066}.
\bibitem[{Montiel et~al.(2021)Montiel, Halford, Mastelini, Bolmier, Sourty,
  Vaysse, Zouitine, Gomes, Read, Abdessalem et~al.}]{riverOl}
\bibinfo{author}{Montiel, J.}, \bibinfo{author}{Halford, M.},
  \bibinfo{author}{Mastelini, S.M.}, \bibinfo{author}{Bolmier, G.},
  \bibinfo{author}{Sourty, R.}, \bibinfo{author}{Vaysse, R.},
  \bibinfo{author}{Zouitine, A.}, \bibinfo{author}{Gomes, H.M.},
  \bibinfo{author}{Read, J.}, \bibinfo{author}{Abdessalem, T.}, et~al.,
  \bibinfo{year}{2021}.
\newblock \bibinfo{title}{River: machine learning for streaming data in python}
  .
\bibitem[{Muzaffar et~al.(2022)Muzaffar, Hassen, Lones and
  Zantout}]{Muzaffar2022}
\bibinfo{author}{Muzaffar, A.}, \bibinfo{author}{Hassen, H.R.},
  \bibinfo{author}{Lones, M.A.}, \bibinfo{author}{Zantout, H.},
  \bibinfo{year}{2022}.
\newblock \bibinfo{title}{An in-depth review of machine learning based android
  malware detection}.
\newblock \bibinfo{journal}{Computers \& Security} , \bibinfo{pages}{102833}.
\bibitem[{Muzaffar et~al.(2023a)Muzaffar, Hassen, Zantout and
  Lones}]{Muzaffar2023}
\bibinfo{author}{Muzaffar, A.}, \bibinfo{author}{Hassen, H.R.},
  \bibinfo{author}{Zantout, H.}, \bibinfo{author}{Lones, M.A.},
  \bibinfo{year}{2023}a.
\newblock \bibinfo{title}{A comprehensive investigation of feature and model
  importance in android malware detection}.
\newblock \URLprefix \url{https://arxiv.org/abs/2301.12778},
  \DOIprefix\doi{10.48550/ARXIV.2301.12778}.
\bibitem[{Muzaffar et~al.(2023b)Muzaffar, Hassen, Zantout and
  Lones}]{muzaffar2023droiddissector}
\bibinfo{author}{Muzaffar, A.}, \bibinfo{author}{Hassen, H.R.},
  \bibinfo{author}{Zantout, H.}, \bibinfo{author}{Lones, M.A.},
  \bibinfo{year}{2023}b.
\newblock \bibinfo{title}{Droiddissector: A static and dynamic analysis tool
  for android malware detection}.
\newblock \href{http://arxiv.org/abs/2308.04170}{{\tt arXiv:2308.04170}}.
\bibitem[{Muzaffar et~al.(2021)Muzaffar, Ragab~Hassen, Lones and
  Zantout}]{Muzaffar2021}
\bibinfo{author}{Muzaffar, A.}, \bibinfo{author}{Ragab~Hassen, H.},
  \bibinfo{author}{Lones, M.A.}, \bibinfo{author}{Zantout, H.},
  \bibinfo{year}{2021}.
\newblock \bibinfo{title}{Android malware detection using api calls: A
  comparison of feature selection and machine learning models}, in:
  \bibinfo{booktitle}{International Conference on Applied CyberSecurity},
  \bibinfo{organization}{Springer}. pp. \bibinfo{pages}{3--12}.
\bibitem[{Narayanan et~al.(2017)Narayanan, Chandramohan, Chen and
  Liu}]{Narayanan2017}
\bibinfo{author}{Narayanan, A.}, \bibinfo{author}{Chandramohan, M.},
  \bibinfo{author}{Chen, L.}, \bibinfo{author}{Liu, Y.}, \bibinfo{year}{2017}.
\newblock \bibinfo{title}{{Context-Aware, Adaptive, and Scalable Android
  Malware Detection Through Online Learning}}.
\newblock \bibinfo{journal}{IEEE Transactions on Emerging Topics in
  Computational Intelligence} \bibinfo{volume}{1}, \bibinfo{pages}{157--175}.
\newblock \DOIprefix\doi{10.1109/tetci.2017.2699220},
  \href{http://arxiv.org/abs/1706.00947}{{\tt arXiv:1706.00947}}.
\bibitem[{Narayanan et~al.(2016)Narayanan, Yang, Chen and
  Jinliang}]{Narayanan2016}
\bibinfo{author}{Narayanan, A.}, \bibinfo{author}{Yang, L.},
  \bibinfo{author}{Chen, L.}, \bibinfo{author}{Jinliang, L.},
  \bibinfo{year}{2016}.
\newblock \bibinfo{title}{{Adaptive and scalable Android malware detection
  through online learning}}.
\newblock \bibinfo{journal}{Proceedings of the International Joint Conference
  on Neural Networks} \bibinfo{volume}{2016-Octob},
  \bibinfo{pages}{2484--2491}.
\newblock \DOIprefix\doi{10.1109/IJCNN.2016.7727508},
  \href{http://arxiv.org/abs/arXiv:1606.07150v2}{{\tt
  arXiv:arXiv:1606.07150v2}}.
\bibitem[{Peiravian and Zhu(2013)}]{Peiravian2013}
\bibinfo{author}{Peiravian, N.}, \bibinfo{author}{Zhu, X.},
  \bibinfo{year}{2013}.
\newblock \bibinfo{title}{{Machine Learning for Android Malware Detection Using
  Permission and API Calls}}, in: \bibinfo{booktitle}{2013 IEEE 25th
  International Conference on Tools with Artificial Intelligence},
  \bibinfo{publisher}{IEEE}. pp. \bibinfo{pages}{300--305}.
\newblock \URLprefix \url{http://ieeexplore.ieee.org/document/6735264/},
  \DOIprefix\doi{10.1109/ICTAI.2013.53}.
\bibitem[{Polychronakis(2017)}]{AMD}
\bibinfo{author}{Polychronakis, Michalis;~Meier, M.}, \bibinfo{year}{2017}.
\newblock \bibinfo{title}{[Lecture Notes in Computer Science] Detection of
  Intrusions and Malware, and Vulnerability Assessment Volume 10327 || Deep
  Ground Truth Analysis of Current Android Malware}. volume
  \bibinfo{volume}{10.1007/978-3-319-60876-1}.
\newblock \URLprefix
  \url{http://gen.lib.rus.ec/scimag/index.php?s=10.1007/978-3-319-60876-1_12},
  \DOIprefix\doi{10.1007/978-3-319-60876-1_12}.
\bibitem[{Rathore et~al.(2021)Rathore, Sahay, Rajvanshi and
  Sewak}]{Rathore2021}
\bibinfo{author}{Rathore, H.}, \bibinfo{author}{Sahay, S.K.},
  \bibinfo{author}{Rajvanshi, R.}, \bibinfo{author}{Sewak, M.},
  \bibinfo{year}{2021}.
\newblock \bibinfo{title}{{Identification of Significant Permissions for
  Efficient Android Malware Detection}}. volume \bibinfo{volume}{355}.
\newblock \bibinfo{publisher}{Springer International Publishing}.
\newblock \URLprefix \url{http://dx.doi.org/10.1007/978-3-030-68737-3{\_}3},
  \DOIprefix\doi{10.1007/978-3-030-68737-3_3}.
\bibitem[{Saracino et~al.(2018)Saracino, Sgandurra, Dini and
  Martinelli}]{Saracino2018}
\bibinfo{author}{Saracino, A.}, \bibinfo{author}{Sgandurra, D.},
  \bibinfo{author}{Dini, G.}, \bibinfo{author}{Martinelli, F.},
  \bibinfo{year}{2018}.
\newblock \bibinfo{title}{{MADAM: Effective and Efficient Behavior-based
  Android Malware Detection and Prevention}}.
\newblock \bibinfo{journal}{IEEE Transactions on Dependable and Secure
  Computing} \bibinfo{volume}{15}, \bibinfo{pages}{83--97}.
\newblock \DOIprefix\doi{10.1109/TDSC.2016.2536605}.
\bibitem[{Shyong et~al.(2020)Shyong, Jeng and Chen}]{Shyong2020}
\bibinfo{author}{Shyong, Y.C.}, \bibinfo{author}{Jeng, T.H.},
  \bibinfo{author}{Chen, Y.M.}, \bibinfo{year}{2020}.
\newblock \bibinfo{title}{{Combining Static Permissions and Dynamic Packet
  Analysis to Improve Android Malware Detection}} ,
  \bibinfo{pages}{75--81}\DOIprefix\doi{10.1109/iccci49374.2020.9145994}.
\bibitem[{Singh et~al.(2012)Singh, Walenstein and Lakhotia}]{singh2012}
\bibinfo{author}{Singh, A.}, \bibinfo{author}{Walenstein, A.},
  \bibinfo{author}{Lakhotia, A.}, \bibinfo{year}{2012}.
\newblock \bibinfo{title}{Tracking concept drift in malware families}, in:
  \bibinfo{booktitle}{Proceedings of the 5th ACM workshop on Security and
  artificial intelligence}, pp. \bibinfo{pages}{81--92}.
\bibitem[{StatCounter(2023)}]{androidData}
\bibinfo{author}{StatCounter}, \bibinfo{year}{2023}.
\newblock \bibinfo{title}{Mobile operating system market share worldwide}.
\newblock \URLprefix
  \url{http://gs.statcounter.com/os-market-share/mobile/worldwide}.
  \bibinfo{note}{[Online; Accessed: January 16, 2023]}.
\bibitem[{Vinod et~al.(2019)Vinod, Zemmari and Conti}]{Vinod2019}
\bibinfo{author}{Vinod, P.}, \bibinfo{author}{Zemmari, A.},
  \bibinfo{author}{Conti, M.}, \bibinfo{year}{2019}.
\newblock \bibinfo{title}{{A machine learning based approach to detect
  malicious android apps using discriminant system calls}}.
\newblock \bibinfo{journal}{Future Generation Computer Systems}
  \bibinfo{volume}{94}, \bibinfo{pages}{333--350}.
\newblock \URLprefix \url{https://doi.org/10.1016/j.future.2018.11.021},
  \DOIprefix\doi{10.1016/j.future.2018.11.021}.
\bibitem[{VirusTotal()}]{virustotal}
\bibinfo{author}{VirusTotal}, .
\newblock \bibinfo{title}{Virustotal}.
\newblock \bibinfo{howpublished}{\url{https://www.virustotal.com}}.
\newblock \bibinfo{note}{[Online accessed January 10, 2023]}.
\bibitem[{Wang et~al.(2016)Wang, Chen, Zhang, Yan, Yang, Peng and
  Jia}]{Wang2016}
\bibinfo{author}{Wang, S.}, \bibinfo{author}{Chen, Z.}, \bibinfo{author}{Zhang,
  L.}, \bibinfo{author}{Yan, Q.}, \bibinfo{author}{Yang, B.},
  \bibinfo{author}{Peng, L.}, \bibinfo{author}{Jia, Z.}, \bibinfo{year}{2016}.
\newblock \bibinfo{title}{{TrafficAV: An effective and explainable detection of
  mobile malware behavior using network traffic}}.
\newblock \bibinfo{journal}{2016 IEEE/ACM 24th International Symposium on
  Quality of Service, IWQoS 2016} \DOIprefix\doi{10.1109/IWQoS.2016.7590446}.
\bibitem[{Wang et~al.(2017)Wang, Zhang, Su and Li}]{Wang2017}
\bibinfo{author}{Wang, X.}, \bibinfo{author}{Zhang, D.}, \bibinfo{author}{Su,
  X.}, \bibinfo{author}{Li, W.}, \bibinfo{year}{2017}.
\newblock \bibinfo{title}{{Mlifdect: Android malware detection based on
  parallel machine learning and information fusion}}.
\newblock \bibinfo{journal}{Security and Communication Networks}
  \bibinfo{volume}{2017}.
\newblock \DOIprefix\doi{10.1155/2017/6451260}.
\bibitem[{Xiao and Yang(2019)}]{Xiao2019a}
\bibinfo{author}{Xiao, X.}, \bibinfo{author}{Yang, S.}, \bibinfo{year}{2019}.
\newblock \bibinfo{title}{An image-inspired and cnn-based android malware
  detection approach}, in: \bibinfo{booktitle}{2019 34th IEEE/ACM International
  Conference on Automated Software Engineering (ASE)},
  \bibinfo{organization}{IEEE}. pp. \bibinfo{pages}{1259--1261}.
\bibitem[{Xiao et~al.(2019)Xiao, Zhang, Mercaldo, Hu and Sangaiah}]{Xiao2019}
\bibinfo{author}{Xiao, X.}, \bibinfo{author}{Zhang, S.},
  \bibinfo{author}{Mercaldo, F.}, \bibinfo{author}{Hu, G.},
  \bibinfo{author}{Sangaiah, A.K.}, \bibinfo{year}{2019}.
\newblock \bibinfo{title}{{Android malware detection based on system call
  sequences and LSTM}}.
\newblock \bibinfo{journal}{Multimedia Tools and Applications}
  \bibinfo{volume}{78}, \bibinfo{pages}{3979--3999}.
\newblock \DOIprefix\doi{10.1007/s11042-017-5104-0}.
\bibitem[{Xu et~al.(2019)Xu, Li, Deng, Chen and Xu}]{Xu2019}
\bibinfo{author}{Xu, K.}, \bibinfo{author}{Li, Y.}, \bibinfo{author}{Deng, R.},
  \bibinfo{author}{Chen, K.}, \bibinfo{author}{Xu, J.}, \bibinfo{year}{2019}.
\newblock \bibinfo{title}{Droidevolver: Self-evolving android malware detection
  system}, in: \bibinfo{booktitle}{2019 IEEE European Symposium on Security and
  Privacy (EuroS\&P)}, \bibinfo{organization}{IEEE}. pp.
  \bibinfo{pages}{47--62}.
\bibitem[{Zulkifli et~al.(2018)Zulkifli, Hamid, Shah and
  Abdullah}]{Zulkifli2018}
\bibinfo{author}{Zulkifli, A.}, \bibinfo{author}{Hamid, I.R.A.},
  \bibinfo{author}{Shah, W.M.}, \bibinfo{author}{Abdullah, Z.},
  \bibinfo{year}{2018}.
\newblock \bibinfo{title}{{Android malware detection based on network traffic
  using decision tree algorithm}}.
\newblock \bibinfo{journal}{Advances in Intelligent Systems and Computing}
  \bibinfo{volume}{700}, \bibinfo{pages}{485--494}.
\newblock \DOIprefix\doi{10.1007/978-3-319-72550-5_46}.

\end{thebibliography}

\end{document}